\pdfoutput=1  % only if pdf/png/jpg images are used

\documentclass[12pt,a4paper]{article}

\usepackage{lineno}  % for line numbering during review
\usepackage{graphicx}  % to include figures (can also use other packages)
\usepackage{xspace} % To avoid problems with missing or double spaces after
\usepackage{color}
\usepackage{colortbl}
\usepackage{amsmath} % Adds a large collection of math symbols
\usepackage{ifthen} % for conditional statements
\usepackage{booktabs}
\newboolean{pdflatex}
\setboolean{pdflatex}{true} % use this if using non-eps figures
\newboolean{articletitles}
\setboolean{articletitles}{true} % False removes titles in references

\newboolean{uprightparticles}
\setboolean{uprightparticles}{false} %Set to true to get roman particle symbols

\usepackage{amssymb}
\usepackage{amsfonts}
\usepackage{upgreek} % Adds in support for greek letters in roman typeset
\usepackage{subfigure}
\usepackage{hyperref}    % Hyperlinks in references
\usepackage[all]{hypcap} % Internal hyperlinks to floats.
\usepackage[font=small]{caption}
\newcommand{\regsim}{\mathord{\sim}}
\newcommand{\hy}{\mathord{-}}

\def\ux85 {UX85\xspace}

{

 \def\PDelta      {\ensuremath{\Delta}\xspace}                 
 \def\PXi      {\ensuremath{\Xi}\xspace}                 
 \def\PLambda      {\ensuremath{\Lambda}\xspace}                 
 \def\PSigma      {\ensuremath{\Sigma}\xspace}                 
 \def\POmega      {\ensuremath{\Omega}\xspace}                 
 \def\PUpsilon      {\ensuremath{\Upsilon}\xspace}                 
 
 \def\PB      {\ensuremath{\mathrm{B}}\xspace}                 
                  
 \def\PD      {\ensuremath{\mathrm{D}}\xspace}

 \def\PK      {\ensuremath{\mathrm{K}}\xspace}

 \def\Pi      {\ensuremath{\mathrm{i}}\xspace}

}
{

 \mathchardef\PDelta="7101
 \mathchardef\PXi="7104
 \mathchardef\PLambda="7103
 \mathchardef\PSigma="7106
 \mathchardef\POmega="710A
 \mathchardef\PUpsilon="7107
                  
 \def\PB      {\ensuremath{B}\xspace}                 
                  
 \def\PD      {\ensuremath{D}\xspace}

 \def\PK      {\ensuremath{K}\xspace}

 \def\Pi      {\ensuremath{i}\xspace}

}

\def\kaon  {\ensuremath{\PK}\xspace}
  \def\Kbar  {\kern 0.2em\overline{\kern -0.2em \PK}{}\xspace}

\def\Kz    {\ensuremath{\kaon^0}\xspace}
\def\Kzb   {\ensuremath{\Kbar^0}\xspace}
\def\KzKzb {\ensuremath{\Kz \kern -0.16em \Kzb}\xspace}
\def\Kp    {\ensuremath{\kaon^+}\xspace}
\def\Km    {\ensuremath{\kaon^-}\xspace}

\def\KpKm  {\ensuremath{\Kp \kern -0.16em \Km}\xspace}

  \def\Dbar    {\kern 0.2em\overline{\kern -0.2em \PD}{}\xspace}
\def\D       {\ensuremath{\PD}\xspace}

\def\Dz      {\ensuremath{\D^0}\xspace}
\def\Dzb     {\ensuremath{\Dbar^0}\xspace}
\def\DzDzb   {\ensuremath{\Dz {\kern -0.16em \Dzb}}\xspace}
\def\Dp      {\ensuremath{\D^+}\xspace}
\def\Dm      {\ensuremath{\D^-}\xspace}

\def\DpDm    {\ensuremath{\Dp {\kern -0.16em \Dm}}\xspace}

  \def\Bbar    {\kern 0.18em\overline{\kern -0.18em \PB}{}\xspace}

  \def\Y#1S{\ensuremath{\PUpsilon{(#1S)}}\xspace}% no space before {...}!

\def\AT#1     {\ensuremath{A_{\mathrm{T}}^{#1}}\xspace}           % 2

\def\C#1      {\ensuremath{\mathcal{C}_{#1}}\xspace}                       % 9
\def\Cp#1     {\ensuremath{\mathcal{C}_{#1}^{'}}\xspace}                    % 7
\def\Ceff#1   {\ensuremath{\mathcal{C}_{#1}^{\mathrm{(eff)}}}\xspace}        % 9  
\def\Cpeff#1  {\ensuremath{\mathcal{C}_{#1}^{'\mathrm{(eff)}}}\xspace}       % 7
\def\Ope#1    {\ensuremath{\mathcal{O}_{#1}}\xspace}                       % 2
\def\Opep#1   {\ensuremath{\mathcal{O}_{#1}^{'}}\xspace}                    % 7

\newcommand{\tev}{\ensuremath{\mathrm{\,Te\kern -0.1em V}}\xspace}
\newcommand{\gev}{\ensuremath{\mathrm{\,Ge\kern -0.1em V}}\xspace}
\newcommand{\mev}{\ensuremath{\mathrm{\,Me\kern -0.1em V}}\xspace}
\newcommand{\kev}{\ensuremath{\mathrm{\,ke\kern -0.1em V}}\xspace}
\newcommand{\ev}{\ensuremath{\mathrm{\,e\kern -0.1em V}}\xspace}
\newcommand{\gevc}{\ensuremath{{\mathrm{\,Ge\kern -0.1em V\!/}c}}\xspace}
\newcommand{\mevc}{\ensuremath{{\mathrm{\,Me\kern -0.1em V\!/}c}}\xspace}
\newcommand{\gevcc}{\ensuremath{{\mathrm{\,Ge\kern -0.1em V\!/}c^2}}\xspace}
\newcommand{\gevgevcccc}{\ensuremath{{\mathrm{\,Ge\kern -0.1em V^2\!/}c^4}}\xspace}
\newcommand{\mevcc}{\ensuremath{{\mathrm{\,Me\kern -0.1em V\!/}c^2}}\xspace}

\def\gsim{{~\raise.15em\hbox{$>$}\kern-.85em
          \lower.35em\hbox{$\sim$}~}\xspace}
\def\lsim{{~\raise.15em\hbox{$<$}\kern-.85em
          \lower.35em\hbox{$\sim$}~}\xspace}

\def\nonn {\ensuremath{\rm {\it{n^+}}\mbox{-}on\mbox{-}{\it{n}}}\xspace}

\def\nonp {\ensuremath{\rm {\it{n^+}}\mbox{-}on\mbox{-}{\it{p}}}\xspace}

\def\tell1  {TELL1\xspace}
\def\ukl1   {UKL1\xspace}

\usepackage{mciteplus}

\begin{document}

% $Id: title-LHCb-PAPER.tex 27192 2012-10-27 22:07:32Z awebber $
% ===============================================================================
% Purpose: LHCb-PAPER journal paper title page template
% Author: 
% Created on: 2010-09-25
% ===============================================================================

%%%%%%%%%%%%%%%%%%%%%%%%%
%%%%%  TITLE PAGE  %%%%%%
%%%%%%%%%%%%%%%%%%%%%%%%%
\begin{titlepage}
\pagenumbering{roman}

% Header ---------------------------------------------------
\vspace*{-1.5cm}
\centerline{\large EUROPEAN ORGANIZATION FOR NUCLEAR RESEARCH (CERN)}
\vspace*{1.5cm}
\hspace*{-0.5cm}
\begin{tabular*}{\linewidth}{lc@{\extracolsep{\fill}}r}
\ifthenelse{\boolean{pdflatex}}% Logo format choice
{\vspace*{-2.7cm}\mbox{\!\!\!\includegraphics[width=.14\textwidth]{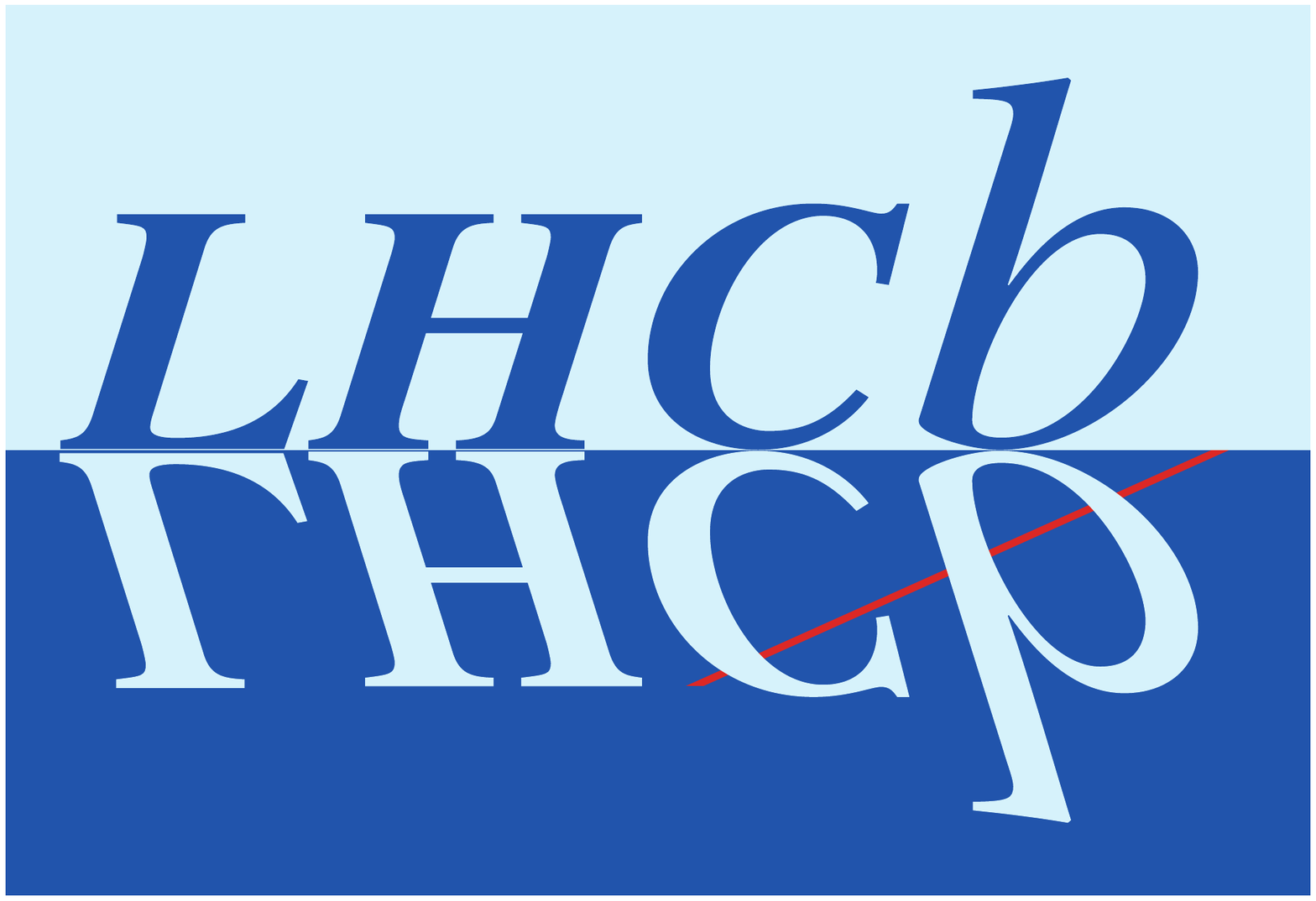}} & &}%
{\vspace*{-1.2cm}\mbox{\!\!\!\includegraphics[width=.12\textwidth]{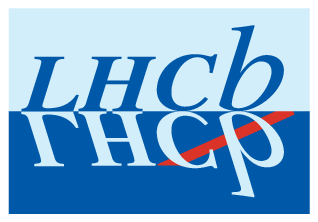}} & &}%
\\
% & & CERN-PH-EP-2012-YYY \\  % ID 
 & & CERN-LHCb-DP-2012-005 \\  % ID 
 & & 21 February 2013 \\%\today \\ % Date - Can also hardwire e.g.: 23 March 2010
 & & \\
% not in paper \hline
\end{tabular*}

\vspace*{2.0cm}

% Title --------------------------------------------------
{\bf\boldmath\huge
\begin{center}
  Radiation damage in the LHCb Vertex Locator
\end{center}
}

\vspace*{1.0cm}

% Authors -------------------------------------------------
\begin{center}
The LHCb VELO group\footnote{Authors are listed on the following pages.}
\end{center}

%\vspace{\fill}

% Abstract -----------------------------------------------
%\newpage
\begin{abstract}
\noindent
The LHCb Vertex Locator (VELO) is a silicon strip detector designed to
reconstruct charged particle trajectories and vertices produced at the LHCb
interaction region. During the first two years of data collection, the $84$ VELO
sensors have been exposed to a range of fluences up to a maximum value of
approximately $\rm{45 \times 10^{12}\,1\,MeV}$ neutron equivalent
($\rm{1\,MeV\,n_{eq}}$). At the operational sensor temperature of approximately
$-7\,^{\circ}\rm{C}$, the average rate of sensor current increase is
$18\,\upmu\rm{A}$ per $\rm{fb^{-1}}$, in excellent agreement with
predictions. The silicon effective bandgap has been determined using current
versus temperature scan data after irradiation, with an average value of
$E_{g}=1.16\pm0.03\pm0.04\,\rm{eV}$ obtained. The first observation of
\nonn{} sensor type inversion at the LHC has been made, occurring at a fluence
of around $15 \times 10 ^{12}$ of $1\,\rm{MeV\,n_{eq}}$. The only \nonp{}
sensors in use at the LHC have also been studied. With an initial fluence of
approximately $\rm{3 \times 10^{12}\,1\,MeV\,n_{eq}}$, a decrease in the
Effective Depletion Voltage (EDV) of around $25$\,V is observed, attributed to
oxygen induced removal of boron interstitial sites. Following this initial
decrease, the EDV increases at a comparable rate to the type
inverted \nonn{} type sensors, with rates of $(1.43\pm 0.16) \times 10
^{-12}\,\rm{V} / \, 1 \, \rm{MeV\,n_{eq}}$ and $(1.35\pm 0.25) \times 10
^{-12}\,\rm{V} / \, 1 \, \rm{MeV\,n_{eq}}$ measured for \nonp{} and
\nonn{} type sensors, respectively. A reduction in the charge collection
efficiency due to an unexpected effect involving the second metal layer readout
lines is observed.
\end{abstract}

\vspace*{4.0cm}
\vspace{\fill}

\end{titlepage}

%This note
%present results from radiation damage studies carried out during the first two
%years of data taking at the LHC. 
%Particular attention has been
%given to the two \nonp{} sensors as this type of sensor is one of the leading
%candidates for the LHC silicon detector upgrades. 

%%%%%%%%%%%%%%%%%%%%%%%%%%%%%%%%
%%%%%  EOD OF TITLE PAGE  %%%%%%
%%%%%%%%%%%%%%%%%%%%%%%%%%%%%%%%

%  empty page follows the title page ----
%\newpage
%\pagenumbering{roman}
\setcounter{page}{3}
%\mbox{~}
%\newpage

% Author List ----------------------------
%  You need to get a new author list!
\begin{center}
\textbf{\large LHCb VELO group}
\end{center}
\begin{flushleft}
%-- Velo Author List - radDamPaperAuthors3012
A.~Affolder$^{1}$,
K.~Akiba$^{2}$
M.~Alexander$^{3}$,
S.~Ali$^{4}$,
M.~Artuso$^{5}$,
J.~Benton$^{6}$,
M.~van Beuzekom$^{4}$,
P.M.~Bj\o rnstad$^{7}$,
G.~Bogdanova$^{8}$,
S.~Borghi$^{3,7}$,
T.J.V.~Bowcock$^{1}$,
H.~Brown$^{1}$,
J.~Buytaert$^{9}$,
G.~Casse$^{1}$,
P.~Collins$^{9}$,
S.~De Capua$^{7}$,
D.~Dossett$^{10}$,
L.~Eklund$^{3}$,
C.~Farinelli$^{4}$,
J.~Garofoli$^{5}$,
M.~Gersabeck$^{9}$,
T.~Gershon$^{9,10}$,
H.~Gordon$^{11}$,
J.~Harrison$^{7}$,
V.~Heijne$^{4}$,
K.~Hennessy$^{1}$,
D.~Hutchcroft$^{1}$,
E.~Jans$^{4}$,
M.~John$^{11}$,
T.~Ketel$^{4}$,
G.~Lafferty$^{7}$,
T.~Latham$^{10}$,
A.~Leflat$^{8,9}$,
M.~Liles$^{1}$,
D.~Moran$^{7}$,
I.~Mous$^{4}$,
A.~Oblakowska-Mucha$^{12}$,
C.~Parkes$^{7}$,
G.D.~Patel$^{1}$,
S.~Redford$^{11}$,
M.M.~Reid$^{10}$,
K.~Rinnert$^{1}$,
E.~Rodrigues$^{3,7}$,
M.~Schiller$^{4}$,
T.~Szumlak$^{12}$,
C.~Thomas$^{11}$,
J.~Velthuis$^{6}$,
V.~Volkov$^{8}$,
A.D.~Webber$^{7}$,
M.~Whitehead$^{10}$,
E.~Zverev$^{8}$.

\bigskip

\footnotesize

\it{
$^{1}$Oliver Lodge Laboratory, University of Liverpool, Liverpool, United Kingdom
\newline$^{2}$Universidade Federal do Rio de Janeiro (UFRJ), Rio de Janeiro,
Brasil 
\newline$^{3}$School of Physics and Astronomy, University of Glasgow, Glasgow, United Kingdom
\newline$^{4}$Nikhef National Institute for Subatomic Physics, Amsterdam, Netherlands
\newline$^{5}$Syracuse University, Syracuse, NY, United States
\newline$^{6}$H.H. Wills Physics Laboratory, University of Bristol, Bristol, United Kingdom
\newline$^{7}$School of Physics and Astronomy, University of Manchester, Manchester, United Kingdom
\newline$^{8}$Institute of Nuclear Physics, Moscow State University (SINP MSU), Moscow, Russia
\newline$^{9}$European Organization for Nuclear Research (CERN), Geneva, Switzerland
\newline$^{10}$Department of Physics, University of Warwick, Coventry, United Kingdom 
\newline$^{11}$Department of Physics, University of Oxford, Oxford, United Kingdom 
\newline$^{12}$AGH University of Science and Technology, Krakow, Poland 
\newline}
\end{flushleft}
%-- Number of authors: 60
%-- Number of institutes: 12

\cleardoublepage

\pagestyle{plain} % restore page numbers for the main text
\setcounter{page}{1}
\pagenumbering{arabic}

\section{Introduction}
\label{sec:Introduction}
The VErtex LOcator (VELO) is a silicon strip detector positioned around the
proton-proton interaction region at the LHCb\,\cite{LHCbDet} experiment. To
obtain the precision vertexing required for heavy-flavour physics, the closest
active silicon sensor region is located $\rm{8.2\,mm}$ from the beam axis,
while the silicon edge is located at a distance of $7$\,mm. For the luminosity
delivered by the LHC in $2010$ and $2011$, the VELO was exposed to higher
particle fluences than any other silicon detector at the LHC.
Careful monitoring of radiation damage to the sensors is essential
to ensure the quality of data for LHCb physics analyses and to provide
information relevant to the eventual detector replacement and upgrade.

\begin{figure}[t]
\centering
\includegraphics[width=0.45\textwidth]{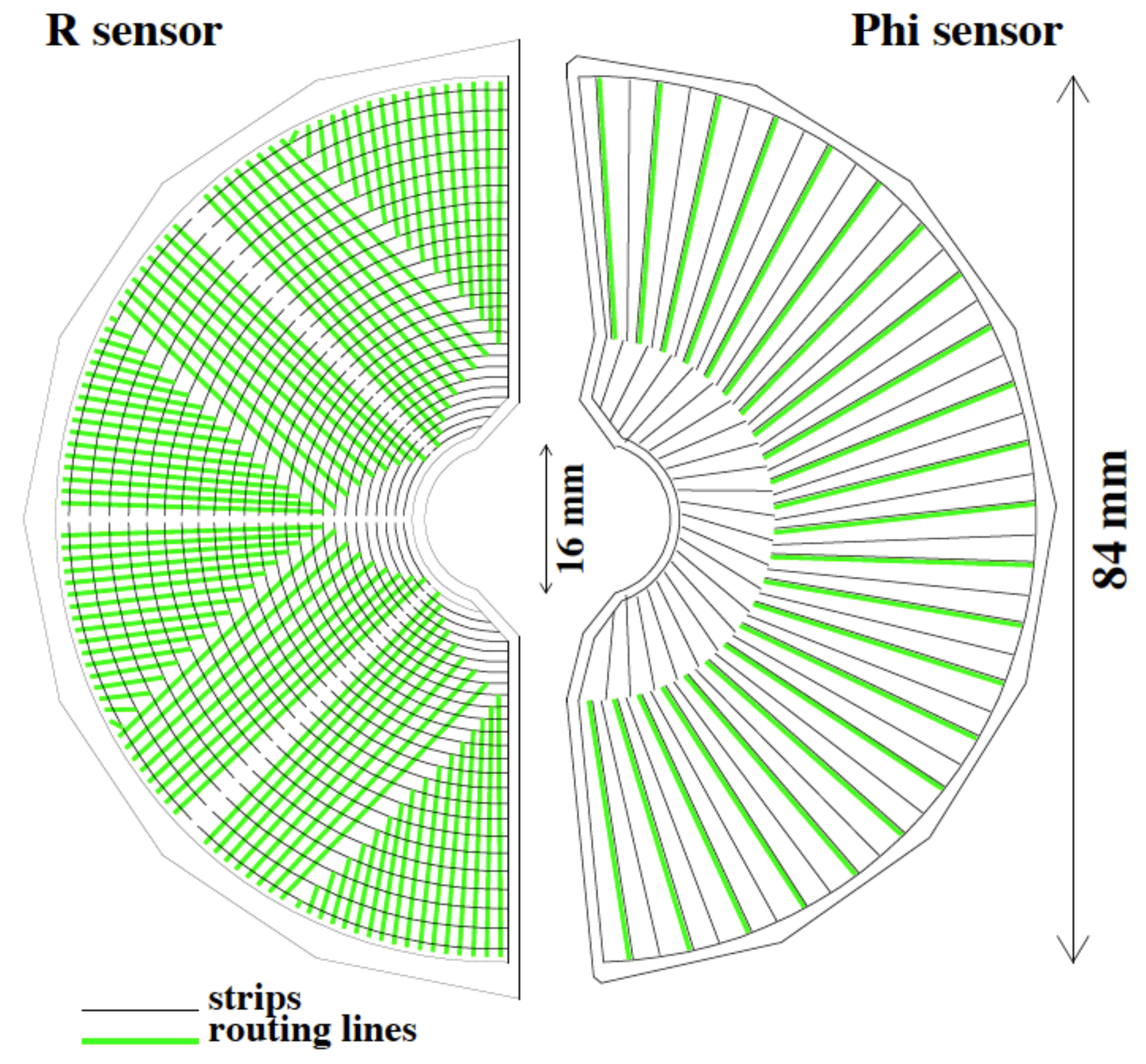}
\caption{A schematic representation of an R-type and a
$\rm{\phi}$-type sensor, with the routing lines orientated perpendicular
and parallel to the silicon strips, respectively.}
\label{fig:VELOSensors}
\end{figure}

During proton injection and energy ramping the LHC beams are wider and less
stable than the beams used for data taking. To prevent damage to the
silicon sensors, the VELO consists of two halves retractable by
$29\,\rm{mm}$ in the horizontal plane. Each half contains $42$ half-disc shaped
silicon-strip sensors. When the beams are in a stable orbit, the two VELO halves
are closed such that the colliding beams are surrounded by the silicon sensors.
Half of the sensors have strips orientated in an approximately radial direction
($\rm{\phi}$-type) and the other half perpendicular to this (R-type), as shown in
figure~\ref{fig:VELOSensors}. A detector module consists of an R-type and
a $\rm{\phi}$-type sensor glued to a common support in a back-to-back
configuration. 
Track coordinates are measured using the charge collected by the two sensors in a module. All but two of the
VELO sensors are oxygenated \nonn{} sensors, consisting of an $n$-type implant
on a $n$-type bulk with a backplane $p^{+}$-type implant. Two oxygenated
\nonp{} silicon sensors are installed at one end of the VELO, intended to be a
test of one of the leading LHC silicon-upgrade candidates in an operational
environment. A summary of the silicon sensor properties is given in
table~\ref{tab:properties}.

Each $n^{+}$ implant is read out via a capacitively coupled \emph{first metal
layer} running along its length. The R-type strips and inner $\rm{\phi}$-type
strips do not extend to the outer region of the sensor. Therefore each strip is
connected via a metal routing line to the edge of the sensor where the readout
electronics are located. The routing lines are referred to as the \emph{second
metal layer} and are insulated from the bulk silicon and first metal layer by
$3.8 \pm 0.3\,\upmu \rm{m}$ of $\rm{SiO_{2}}$. For R-type sensors, the routing lines
are positioned perpendicular to the sensor strips, whilst for the
$\rm{\phi}$-type they are positioned directly above and parallel to the strips.

\begin{table}[b]
\centering
\small
\caption{The VELO sensor design parameters. The sensor position along the beam
axis is given relative to the beam interaction region.}
\begin{tabular}{@{}lr@{}}
\toprule
Parameter&Value\\
\midrule
Silicon thickness&$300\,\upmu \rm{m}$\\
Strip pitch&$40 \hy 120\,\upmu \rm{m}$\\
Strip width & $ 11 \hy 38\,\upmu \rm{m}$ \\
Routing line width & $\regsim 11\,\upmu \rm{m}$ \\
Inner silicon edge to beam axis&$7\,\rm{mm}$\\ 
Radial distance of active strips from beam axis&$8.2 \hy 42\,\rm{mm}$\\ 
Sensor position along beam-axis&$-300~\rm{to}~750\,\rm{mm}$\\ 
Oxygen enhancement&$>1\,\rm{x}\,10^{17}\,\rm{cm^{-3}}$\\
\bottomrule
\end{tabular}
\label{tab:properties}
\end{table}

This paper presents studies of radiation damage in the VELO sensors using data
collected during $2010$ and $2011$. Section\,\ref{sec:currents} describes
studies of sensor currents as a function of voltage and temperature. In
section~\ref{sec:depvoltstudies} the effects of radiation are monitored by
measuring the charge collection efficiency and noise as a function of bias
voltage. An unexpected decrease in clustering efficiency due to an effect
involving the second metal layer is described in
section~\ref{sec:secondmetallayer}. The results from the various studies are
summarised in section~\ref{sec:Summary}.

 \section{Current evolution}
\label{sec:currents}
The leakage current in a silicon sensor varies linearly with fluence for a wide
range of silicon substrates\,\cite{Moll}. This predictability provides a simple
and accurate method of relating sensor currents to the amount of particle
fluence accumulated by a sensor. This section presents an analysis of sensor
currents in order to monitor radiation damage in the VELO.

\begin{figure}[b]
\centering
\includegraphics[width=0.7\textwidth]{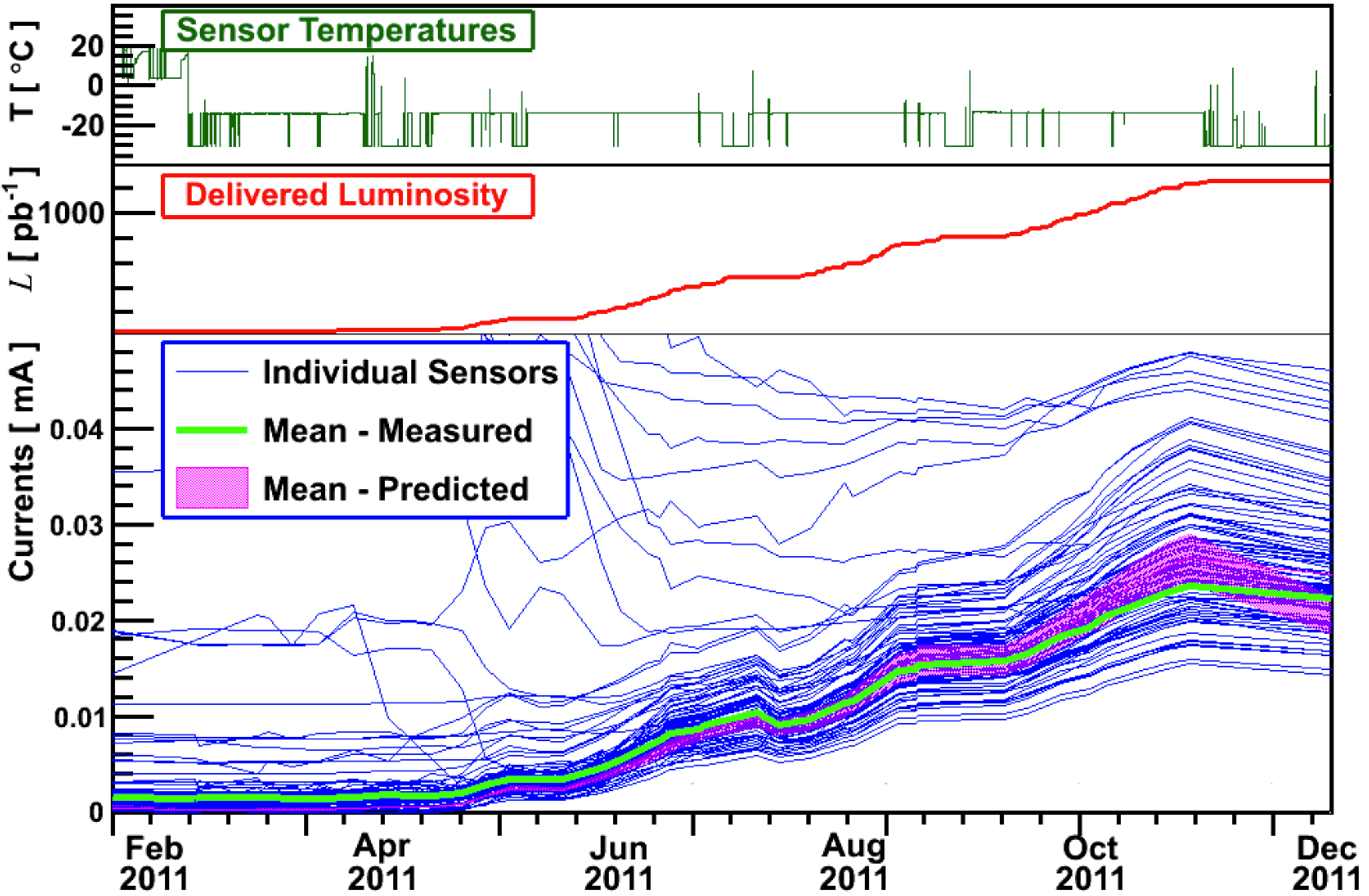}
\caption{Currents measured in the VELO for each sensor as
a function of time (bottom). The luminosity delivered to LHCb and the average
sensor temperature is shown over the same time scale (middle and top). Increases
in the delivered luminosity are matched by increases in the sensor currents.
The mean measured current agrees well with the prediction from simulation,
which is described in Sect.2.4.
The mean measured value excludes
sensors that are surface-current dominated.}
\label{fig:CurLumTime}
\end{figure}

\subsection{Interpretation of sensor current measurements}
\label{sec:currentsInt}
Each VELO sensor is biased through a high-voltage connection to the backplane, and
a common ground connection to the bias line and innermost guard ring on the
strip side. Therefore a single current is
measured, corresponding to the current drawn by the entire sensor.
The raw measured currents are shown as a function of time in
figure~\ref{fig:CurLumTime}, for sensors operated at the nominal bias voltage of
$150\,\rm{V}$ and at a mean temperature of approximately $-7\,^{\circ}\rm{C}$.
During the early running period, the majority of the sensors had small currents,
with a few exceptions that had currents of up to $40\,\upmu \rm{A}$. With
sensor irradiation the bulk currents have increased. The spread of the measured
currents is partly due to the difference in the sensor positions relative to
the interaction point, but is dominated by variations in the sensor
temperatures. 
The occasional dips in the currents are related to short annealing periods in which the cooling systems were switched off.

The measured currents contain contributions from two dominant sources,
generically referred to as \emph{bulk} and \emph{surface} currents.
The bulk currents vary exponentially with temperature and have a precisely
predicted relationship with fluence. Surface currents arise due to
irregularities introduced during sensor production such as process errors,
scratches, guard rings and non-uniformities in the cut edges. Some of these
contributions may have an exponential dependence on
temperature\,\cite{TempDepCur}, however the surface currents measured in VELO
sensors are predominantly characterised by an Ohmic increase in current with
bias voltage. In the majority of sensors the Ohmic surface current is seen
to anneal with particle fluence. An analysis of VELO sensor currents as a
function of bias voltage (IV scan) is described in detail in
ref.\,\cite{IVNote}.

\begin{figure}[b]
\centering
\subfigure[]{
  \includegraphics[width=0.45\textwidth]{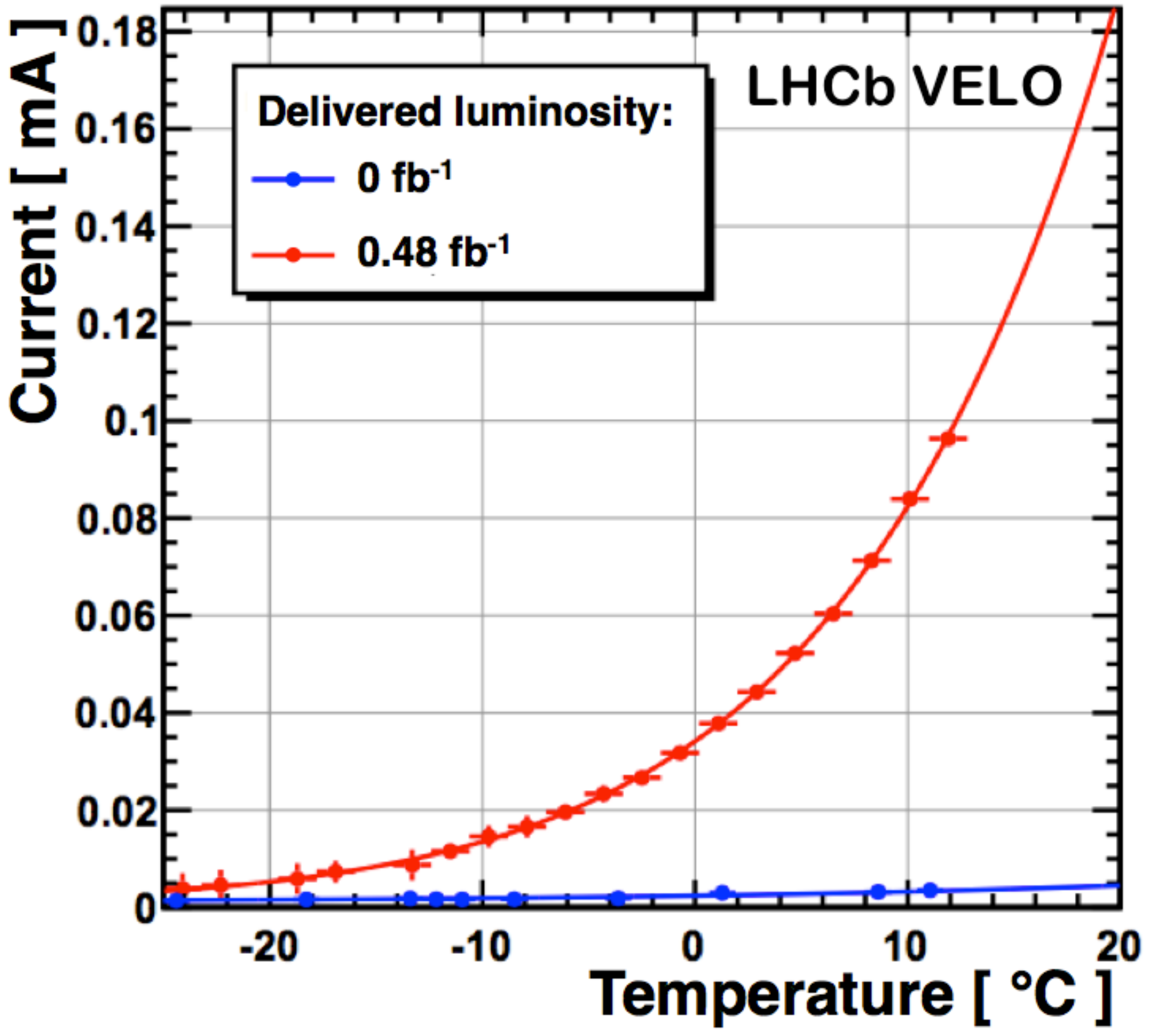}
  \label{fig:Bulk}
}
\subfigure[]{
  \includegraphics[width=0.447\textwidth]{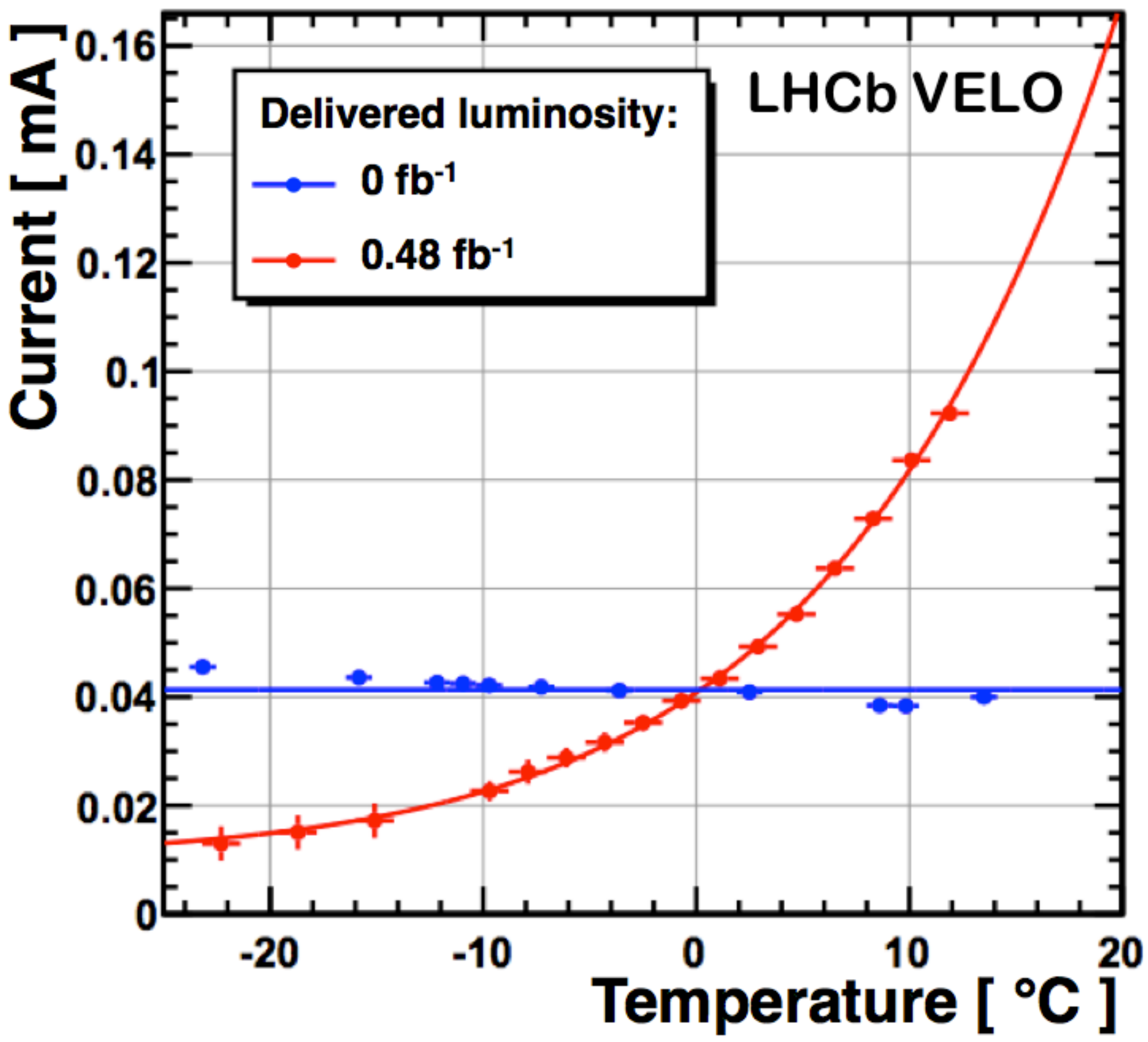}
  \label{fig:Surface}
}
\caption{The current versus temperature for two VELO sensors operated at
the nominal $150\rm{\,V}$ bias. Before irradiation there is a small exponential
component for the sensor shown in figure~\textbf{a)}, while that shown in
figure~\textbf{b)} is dominated by the non-temperature dependent surface current
contribution. After irradiation, a large exponential component is seen for both
sensors. These sensors are now bulk current dominated and the surface current
has largely been annealed.}
\label{fig:CatSens}
\end{figure}

For the VELO sensors the pre-irradiation exponential contribution is very small
and is assumed to consist of a mixture of bulk and surface currents. The
relative contribution of bulk and surface current is identified by measuring
the current as a function of temperature (IT scan), as shown in
figure~\ref{fig:CatSens}. This method is described in detail in
ref.\,\cite{ITNote}.

\subsection{The effective band gap}
\label{sec:EffGap}
In the temperature range of interest, the bulk current is expected to scale
according to,
\begin{equation} 
\label{Equ:CurVsTemp1}
I(T) \propto T^{2}\exp \left(-\frac{E_{g}}{2kT}\right),
\end{equation}
where $T$ is the temperature in Kelvin and $k$ is the Boltzmann constant
($8.62\times10^{-5}\,\rm{eV}$). The constant $E_{g}$ is related to the bandgap
energy and an effective value of $1.21\,\rm{eV}$ is assumed for the temperature range of
interest\,\cite{EgGap}. Using this function the sensor currents can be normalised
to a common temperature. In addition, the current variation as a function of
temperature can be fitted to measure $E_{g}$ in the VELO sensors. A summary of
the results obtained from several IT scans is presented in
table~\ref{tab:EffBand}. The statistical uncertainty quoted is the width
of a Gaussian fitted to the distribution of the measured values from all sensors.
The largest source of systematic uncertainty is related to the accuracy with
which sensor temperatures are known, and the choice of the temperature fitting
range\,\cite{ITNote}.

\begin{table}[t]
\centering
\caption{The effective band gap, $E_{g}$, measured following various amounts of
delivered luminosity. The first uncertainty is statistical and the second is
systematic.}
\begin{tabular}{@{}ccc@{}}
\toprule
Delivered luminosity&Bias voltage&$E_{g}$\\ 
$\rm{[\,fb^{-1}\,]}$&[\,V\,]&[\,eV\,]\\
\midrule
$0.48$&$100$&$1.17 \pm 0.07 \pm 0.04$\\
$0.48$&$150$&$1.18 \pm 0.05 \pm 0.04$\\
$0.82$&$150$&$1.14 \pm 0.06 \pm 0.04$\\
$1.20$&$150$&$1.15 \pm 0.04 \pm 0.04$\\
\bottomrule
\end{tabular}
\label{tab:EffBand}
\end{table}

The weighted average of the measured values is
$E_{g}=1.16\pm0.03\pm0.04\,\rm{eV}$, which is statistically compatible with the
expected value of $1.21\,\rm{eV}$ from the literature\,\cite{EgGap}. Recent
studies\,\cite{EgGap} have shown that discrepancies can occur due to sensor
self-heating, in which case $E_{g}$ is measured to be systematically high. In
addition, dependencies of $E_{g}$ on the sensor bias voltage have been
observed. However, the VELO sensors used for these measurements were cooled and
sufficiently biased at $150$\,V, such that these effects should not
significantly influence the result. This is supported by the consistency of
the cross-check measurement made at $100$\,V.

\subsection{Fluence determination}
\label{FluDet}
With a good understanding of the annealing conditions, changes in current can
be related to the particle fluence incident on a sensor. The expected change in
leakage current at room temperature is given by the relation,
\begin{equation} 
\label{Equ:CurVsTemp}
\Delta I = \alpha \phi V_{Si},
\end{equation}
where $\alpha$ is the annealing parameter in units $\rm{A}\,\rm{cm}^{-1}$,
$\phi$ is the fluence in units of number of particles per $\rm{cm^{2}}$, and $V_{Si}$
is the silicon volume in $\rm{cm^{3}}$. The annealing parameter is a logarithmic
function of time, and also depends on temperature. It has been shown to follow
an Arrhenius relation,
\begin{equation} 
\label{Equ:alpha}
\alpha  \propto \exp \left(-\frac{E_{a}}{kT}\right),
\end{equation}
where $E_{a}$ is the activation energy 
for which a value of $1.33 \pm 0.07 \,\rm{eV}$ has been derived from a fit \,\cite{ActEnergy}. In order to estimate the appropriate
value of $\alpha$ for our data, we proceed as follows. The sensor
temperatures are recorded for each minute of operation and corrected to an
equivalent time at $21\,^{\circ}\rm{C}$. The delivered luminosity,
$\mathcal{L}$, is folded with this equivalent time to produce an effective
value of $\alpha \phi$ to be used in eq.\,\ref{Equ:CurVsTemp}. This
procedure\,\cite{ITNote} yields a value for $\alpha \mathcal{L}$ of $4.8
\times 10^{-17}\,\rm{A\,fb^{-1}\,cm^{-1}}$, corresponding to an effective value
for $\alpha$ of $6 \times 10^{-17}\,\rm{A\,cm^{-1}}$.

Fluences simulated with GEANT4 in the LHCb detector are folded with particle
damage factors using the NIEL scaling hypothesis\,\cite{NIEL}. The damage caused
by high energy particles in the bulk material is parameterised as proportional
to a displacement damage cross section, $D$. A displacement damage cross
section of $95$\,MeV\,mb for $1$\,MeV neutrons, $D_{n}(1\,\rm{MeV})$, is assumed 
and each particle, $i$, is assigned an energy-dependent
hardness factor, $k_{i}$,
\begin{equation} 
\label{Equ:DamageFact}
k_{i}(E)  = D_{i}(E)/D_{n}(1\,\rm{MeV}),
\end{equation}
which can be used to estimate the damage in the material. The total damage is
expressed as a multiple of the expected damage of a $1$\,MeV neutron (referred
to as $1$\,MeV neutron equivalent, or $1$\,MeV\,$\rm{n_{eq}}$). This technique
has been shown to be highly effective for describing the evolution of the leakage
current. For this analysis, every particle in the simulation that is incident on
a VELO sensor is assigned a displacement damage cross-section, or cross-section per
silicon atom, using as reference a set of tabulated values 
from ref.\,\cite{DamFact}.

\begin{figure}[t]
\centering
\subfigure[]{
  \includegraphics[width=0.51\textwidth]{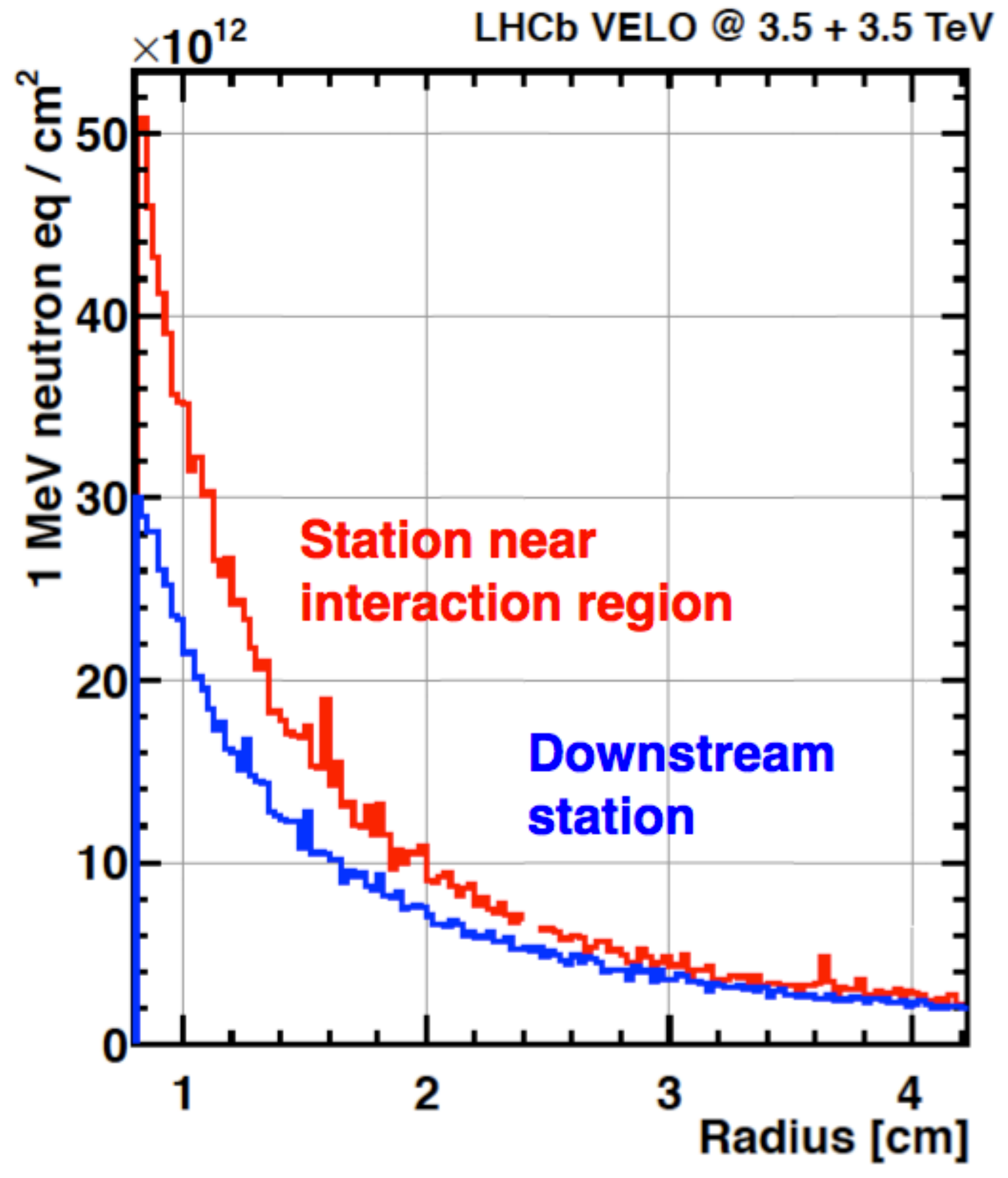}
  \label{fig:currentsplot1A}
}
\subfigure[]{
  \includegraphics[width=0.34\textwidth]{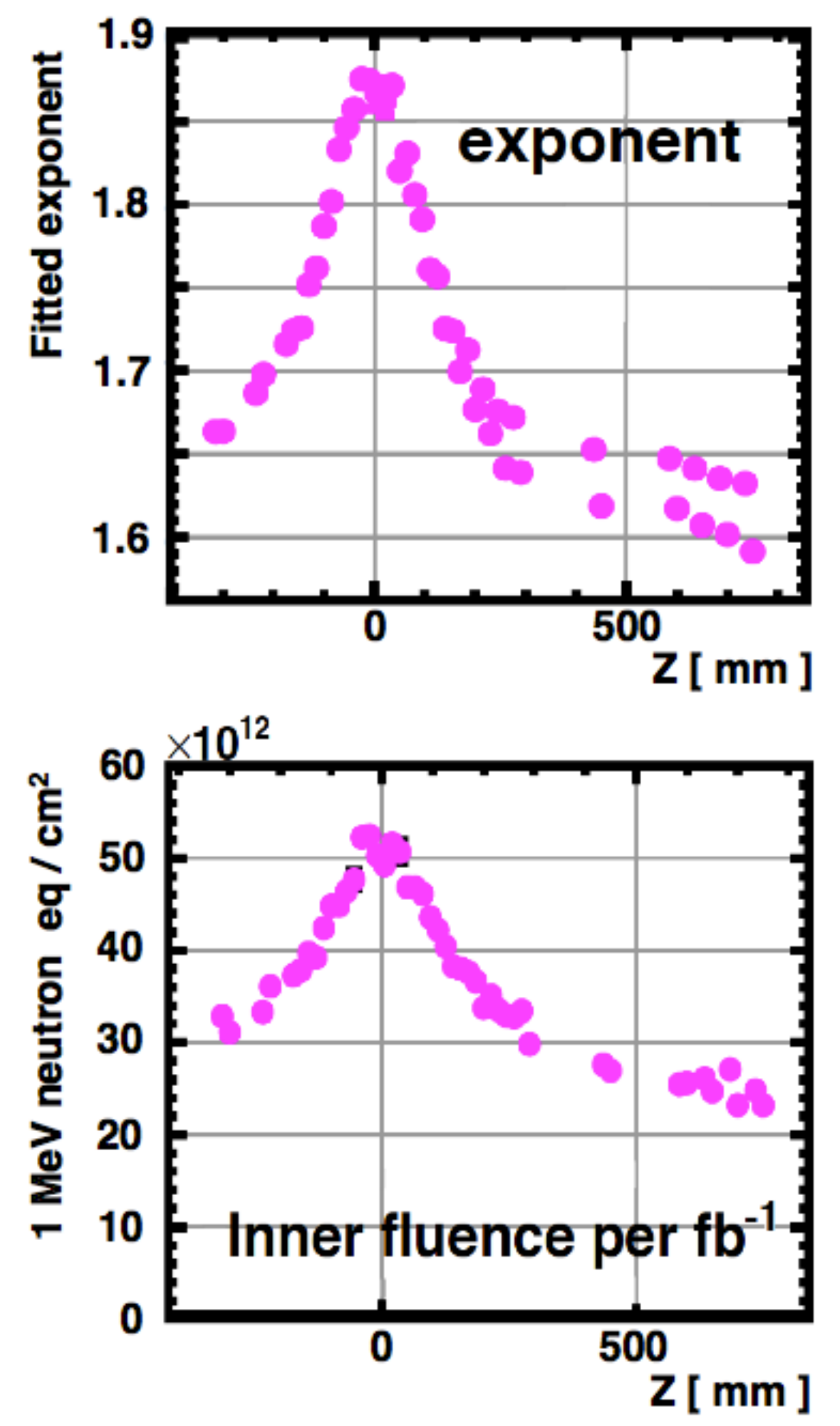}
  \label{fig:currentsplot1B}
}
\caption{\textbf{a)} The fluence from $\rm{1\,fb^{-1}}$ of integrated
luminosity versus radius for two VELO sensors, as seen in simulated
proton-proton collisions at a $7$\,TeV centre-of-mass energy. \textbf{Top b)}
In each sensor, the fluence as a function of radius is fitted with the function
$\rm{Ar^{k}}$. The fitted exponent, ${\rm k}$, is shown as a function of the
sensor $z$-coordinate, where $z$ is the distance along the beam-axis that
a sensor is from the interaction region. The distribution of the fluence
across the sensor is seen to become flatter with increasing distance from the
interaction region. \textbf{Bottom b)} The fluence at the innermost radius of
the sensor against the sensor $z$-coordinate.}
\label{fig:currentsplot1}
\end{figure}

The fluence varies with the geometric location of the sensors in the VELO, as
shown for simulated events in figure~\ref{fig:currentsplot1}. It decreases with
radial distance from the beam with an approximate $1/r^{1.75}$ dependence. The
position along the beam-pipe ($z$-direction) and path length of the particle in
the silicon sensor were both found to significantly affect the fluence
incident on a sensor. The predicted fluence is assigned an uncertainty of $\regsim
8\%$ which is dominated by uncertainties due to particles with no available displacement
damage data. The average measured current increase for the sensors at the operational
temperature of approximately $-7\,^{\circ}\rm{C}$ (and including several
annealing periods) is $\regsim 18\,\upmu$A per $\rm{fb^{-1}}$. 

\subsection{Predicted currents}
\label{sec:Predcur}
\begin{figure}
\centering
\includegraphics[width=0.7\textwidth]{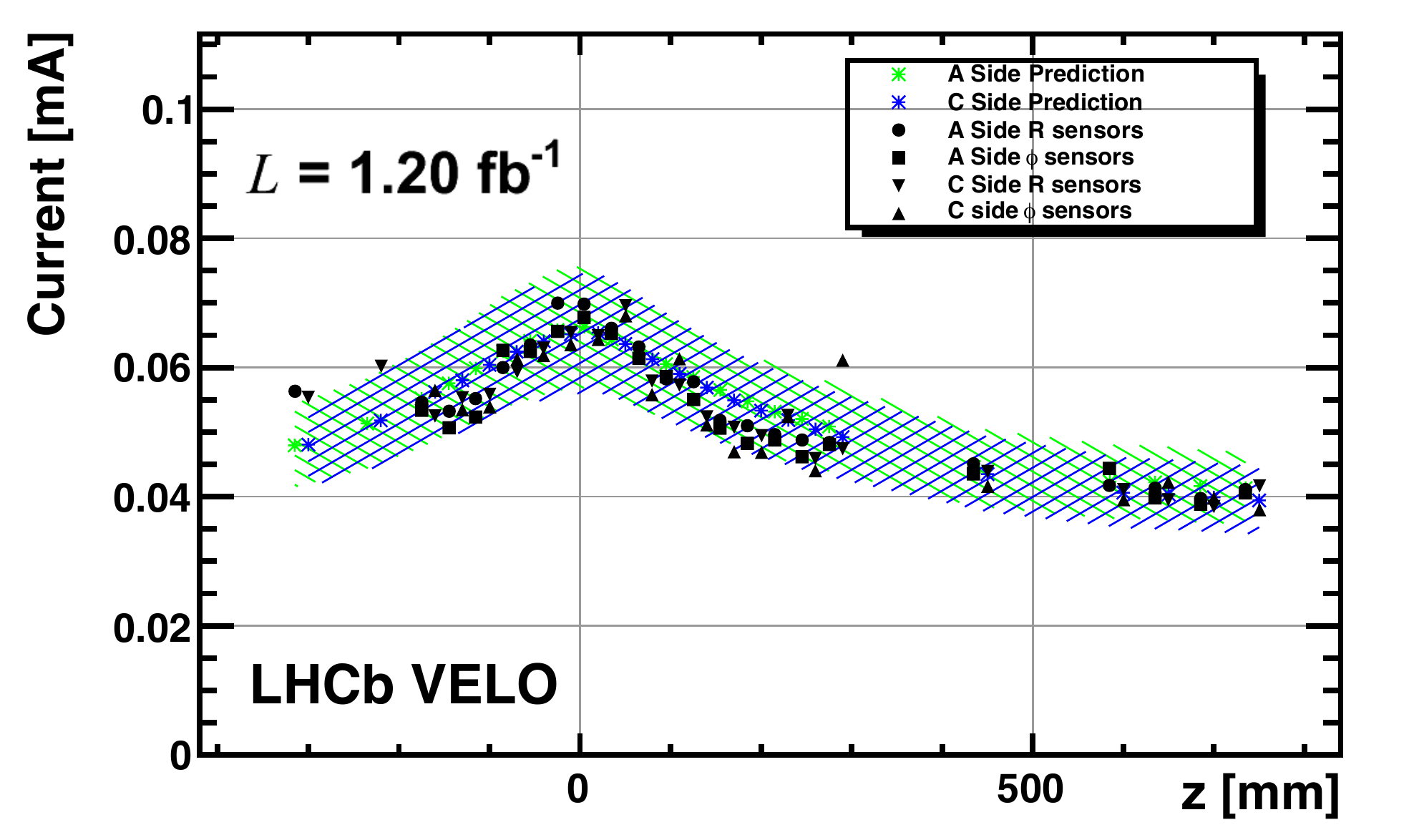}
\caption{The leakage current against sensor $z$-coordinate after
$\rm{1.20\,fb^{-1}}$ of integrated luminosity, normalised to $0\,^{\circ}{\rm
C}$. The data is in agreement with predictions, represented by the shaded
region. The two VELO halves are referred to as the A and C sides of the VELO.}
\label{fig:currentsplot2}
\end{figure}

With the relationship between luminosity and damaging fluence established, the
expected change in leakage current due to radiation damage can be predicted.
The integrated luminosity is taken from the LHCb online measurement.
Predictions are shown to agree with the measured currents, as shown in
figure~\ref{fig:currentsplot2}. The predictions have an associated uncertainty of
approximately $10\%$, estimated by adding the uncertainties for the integrated
luminosity ($5\%$), the annealing factor ($3\%$) and the damaging fluence
prediction ($8\%$) in quadrature. 
The predicted leakage current is on average within $5\%$ of the measured current.

\section{Depletion voltage studies}
\label{sec:depvoltstudies}
The depletion voltage, defined as the reverse bias voltage required to fully
deplete a sensor, has been monitored as a function of the delivered
luminosity. For fluences delivered to the VELO within the first few
years of operation, the change in depletion voltage for an \nonn{} type
sensor should be accurately described by the Hamburg model\,\cite{Hamburg}. The
effective doping of the $n$-bulk changes over time due to radiation-induced
defects. Dominant mechanisms are expected to be the inactivation of
phosphorous dopants in combination with the introduction of acceptors. This
leads to an initial decrease in the depletion voltage of the \nonn{} sensors to
a value close to $\rm{0\,V}$. The $n$-type bulk inverts and becomes
$p$-type, after which further irradiation leads to an increase in depletion
voltage. Eventually, the bias voltage required to obtain sufficient
charge from a sensor will cause electrical breakdown in the silicon or
exceed the $500$\,V hardware limit, thus limiting the useful lifetime of the
silicon detector. For an oxygenated \nonp{} type sensor irradiated with charged
hadrons there are expected to be competing mechanisms, with acceptor
introduction partially compensated by initial oxygen-induced acceptor
removal\,\cite{nonpPred, SensComp}.

Following manufacture the depletion voltage of each VELO sensor was measured by
comparing the capacitance (C) to the bias voltage (V)\,\cite{LiverpoolDV}. It is not possible to implement this technique after
VELO installation and so alternative methods are used to extract information on the depletion voltage. This section presents results
from two such studies: Charge Collection Efficiency and noise scan studies. 

\subsection{Charge Collection Efficiency}
\label{sec:CCE}
The amount of charge collected by an under-depleted silicon strip increases
with the bias voltage applied. When the sensor is fully depleted, any further
increase in bias voltage will not increase the amount of charge
collected (given a sufficient signal collection time). The relationship between
the Charge Collection Efficiency (CCE) and the applied bias voltage has been
exploited to measure a property of the sensor analogous to the depletion
voltage, referred to as the Effective Depletion Voltage (EDV).

\subsubsection{Effective Depletion Voltage determination}
\label{sec:CCEmethod}
\begin{figure}
\centering
\subfigure[]{
  \includegraphics[width=0.47\textwidth]{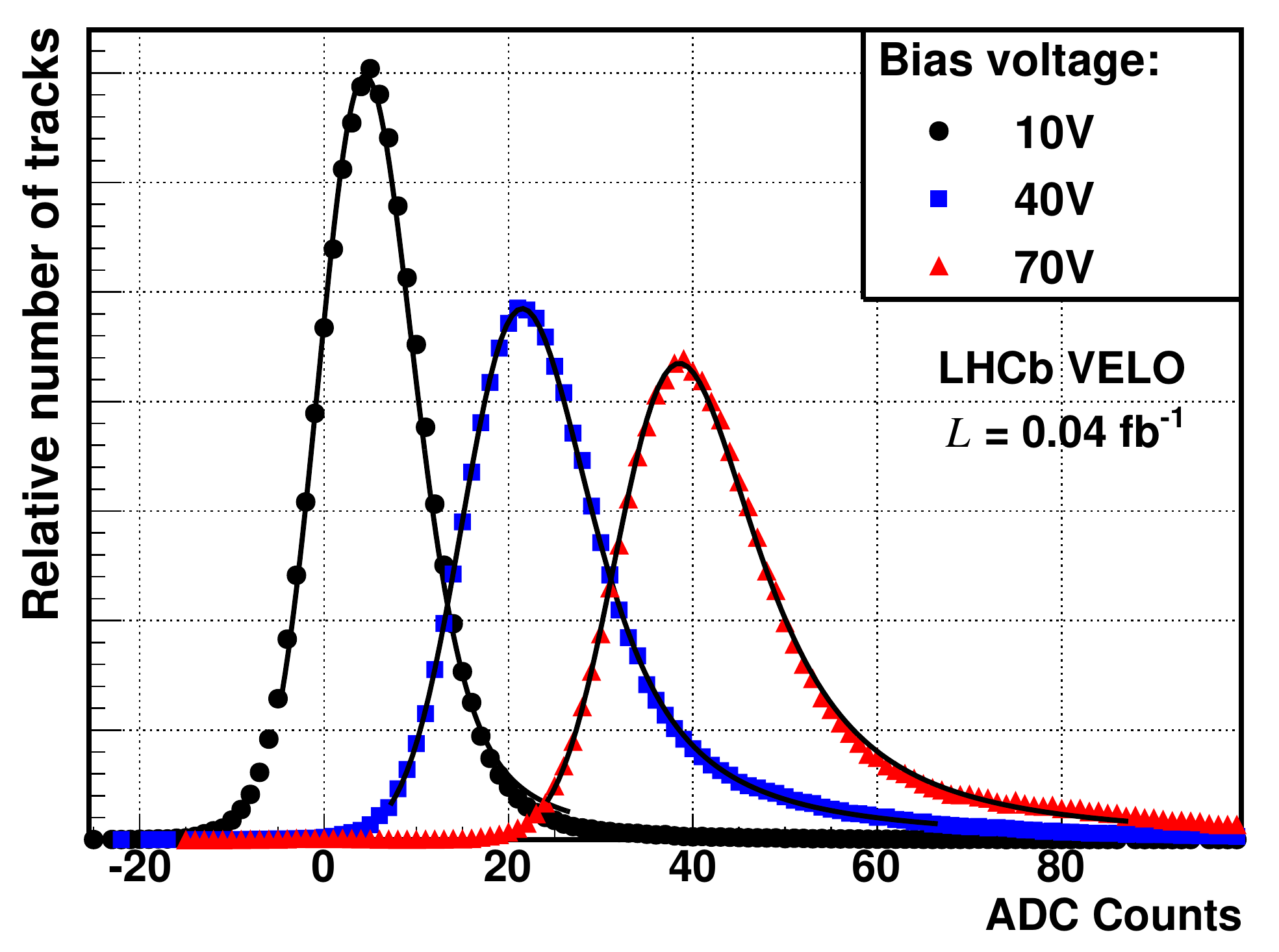}
  \label{fig:EDVmethodA}
 }
\subfigure[]{
  \includegraphics[width=0.47\textwidth]{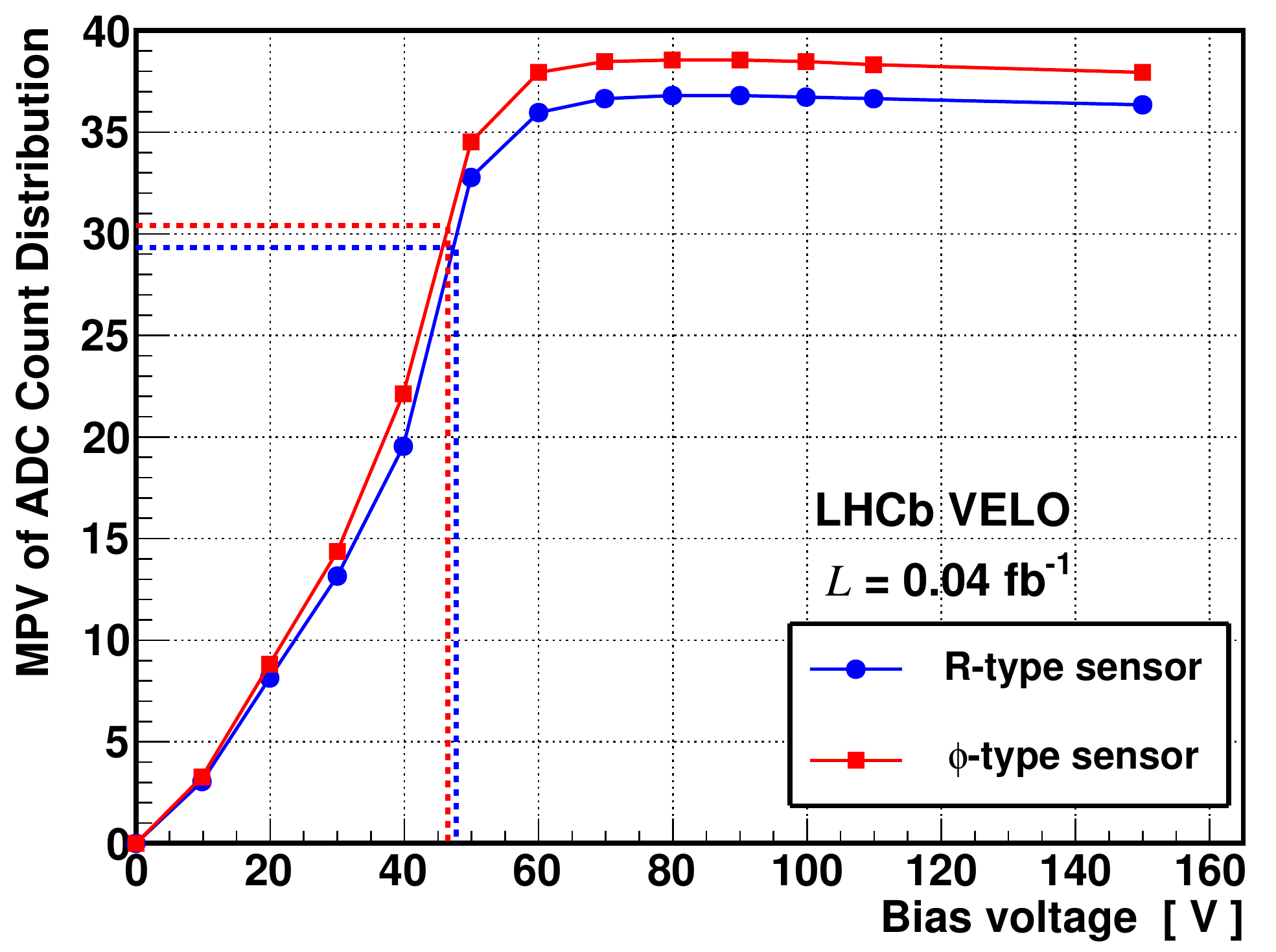}
  \label{fig:EDVmethodB}
}
\caption{\textbf{a)} The pedestal subtracted ADC distributions for an R-type
sensor at three example bias voltages. \textbf{b)} The MPV of the fit to the
ADC distribution vs. bias voltage. The dashed lines represent the ADC that is
$80\%$ of the plateau value, and the corresponding EDV.}
\label{fig:EDVmethod}
\end{figure}

The nominal operational voltage of the VELO sensors is $\rm{150\,V}$. For the
CCE analysis, collision data is recorded with every fifth module operated at a
voltage ranging between $0$ and $\rm{150\,V}$. The remaining modules are
maintained at $\rm{150\,V}$. Sensors with variable voltage are
referred to as \emph{test} sensors. The test sensors are removed from the
reconstruction algorithms such that only hits from the $\rm{150\,V}$ operated
sensors are used to reconstruct particle tracks. A track is extrapolated to a
coordinate on the test sensor and the set of five strips nearest to this
coordinate are searched for deposited charge. This provides unbiased information on the
amount of charge deposited by the particle as a function of bias voltage. At
each bias voltage the pedestal-subtracted ADC distribution is fitted using a
gaussian convoluted with a Landau function. This is used to determine the Most
Probable Value (MPV) of the ADC distribution. At large bias voltages the MPV of
the ADC distribution reaches a plateau. The EDV is defined as the voltage at
which the MPV of a sensor is equal to $80\%$ of the plateau ADC value, as shown in
figure~\ref{fig:EDVmethod}. The threshold of $80\%$ was chosen as it gives
closest agreement with depletion voltages determined from pre-irradiation CV
measurements\,\cite{LiverpoolDV}. For unirradiated sensors, the difference in the
values obtained using the CV method and the EDV method is less than $10\,\rm{V}$.

Differences between the depletion voltage and EDV are expected near to sensor
type-inversion. The depletion voltage is expected to decrease to a value of
approximately $0$\,V, whereas the minimum EDV is dictated by the smallest
potential difference required to collect charge from the silicon strips, which
in turn depends on the shaping time of the electronics. When operated with a
$25\,$ns signal shaping time, the smallest EDV that can be measured in the VELO
sensors is approximately $20$\,V. The EDV is therefore not an accurate estimate
of the depletion voltage below this value.

\subsubsection{Bulk radiation damage}
\label{sec:CCEBulkDam}
\begin{figure}[tbp]
\centering
\subfigure[]{
      \includegraphics[width=0.47\textwidth]{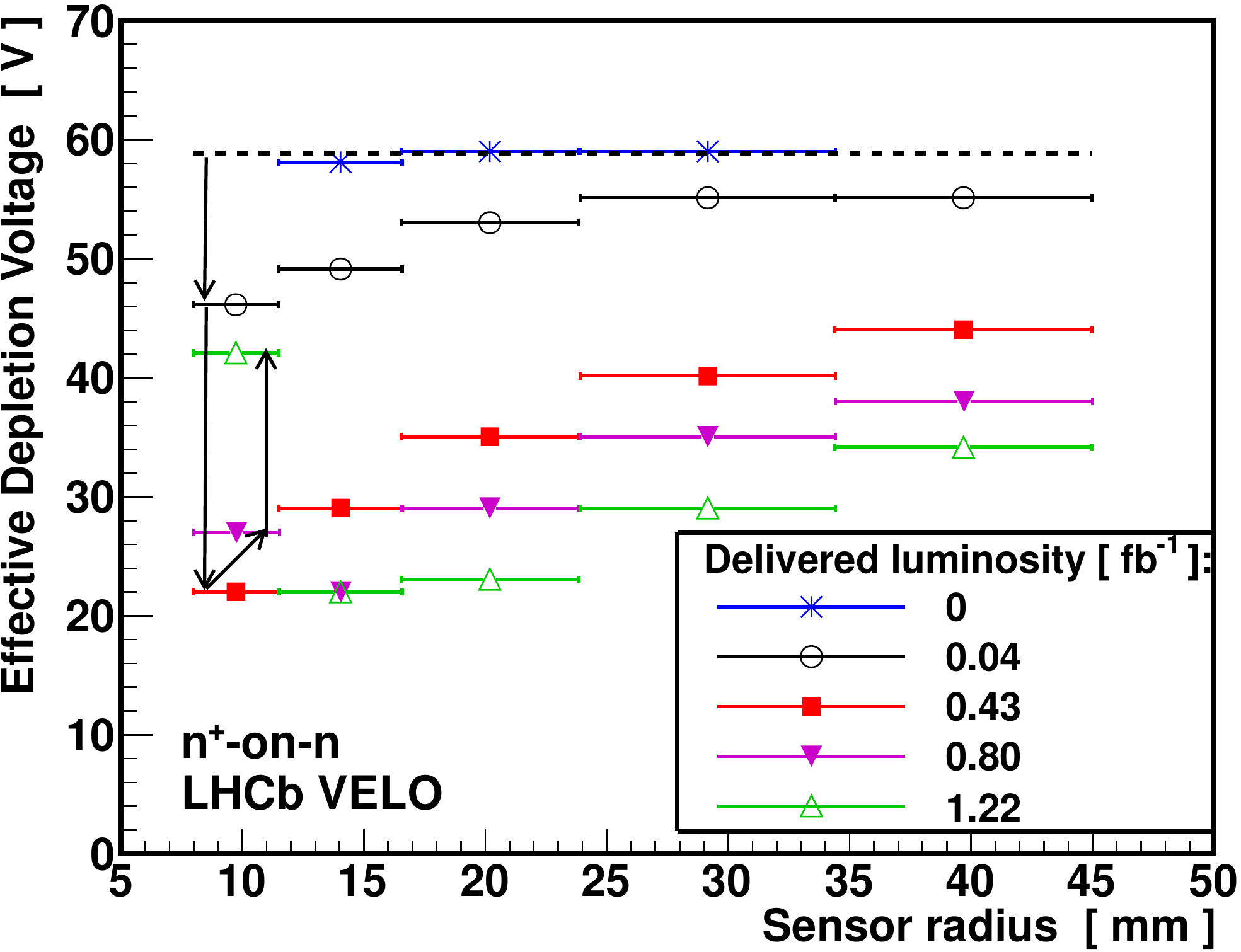}
      \label{fig:EDVDropRadA}
}
\subfigure[]{
      \includegraphics[width=0.47\textwidth]{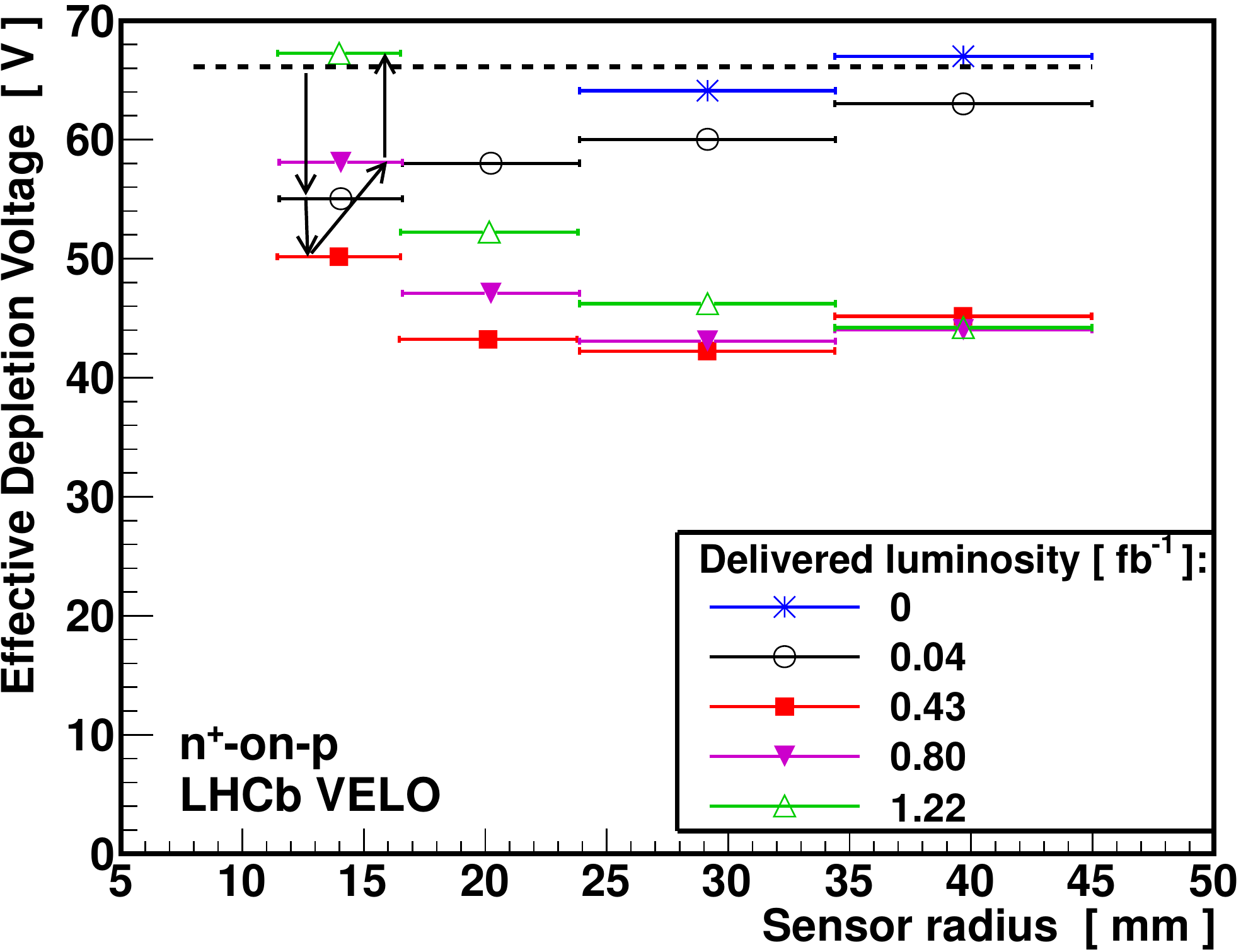}
      \label{fig:EDVDropRadB}
}
\caption{\textbf{a)} The EDV against sensor radius for an \nonn{} type
sensor for each of the CCE scans. The dashed line shows the
mean EDV across all radius regions prior to sensor irradiation, where some
$0\,\rm{fb^{-1}}$ data points are not present due to low statistics. \textbf{b)}
A similar plot for the \nonp{}, $\phi$-type sensor. The minimum EDV is
$\rm{\regsim 40\,V}$, which is significantly higher than the minimum at
$\rm{\regsim 20\,V}$ observed for the \nonn{} type sensor.}
\label{fig:EDVDropRad}
\end{figure}

Between April $2010$ and October $2011$ five dedicated CCE scans were taken,
corresponding to delivered luminosities of $0$, $0.04$, $0.43$, $0.80$ and
$1.22\,\rm{fb^{-1}}$. As the fluence delivered to the sensors varies
significantly with sensor radius, each sensor is divided into $5$ radial
regions such that the fluence does not change by more than a factor of two
across a region. The change in EDV with irradiation for a particular \nonn{}
type sensor is shown in figure~\ref{fig:EDVDropRadA}. Initially the EDV is found to decrease with fluence across all radial regions, 
as predicted by the Hamburg model. The rate of decrease is greater in the inner radial regions of
the sensor, consistent with expectations that these regions are exposed to
higher fluence. The innermost region undergoes an increase in EDV between
$0.80$ and $1.22\,\rm{fb^{-1}}$ of delivered luminosity, indicating that this
part of the sensor has undergone type inversion. 
The \nonp{} type sensors exhibit a decrease in EDV with initial fluence, as
shown in figure~\ref{fig:EDVDropRadB}. This is understood to
be caused by oxygen induced removal of boron interstitial acceptor sites, an
effect that has been previously observed\,\cite{nonpPred, SensComp}.

\begin{figure}[tbp]
\centering
\includegraphics[width=0.58\textwidth]{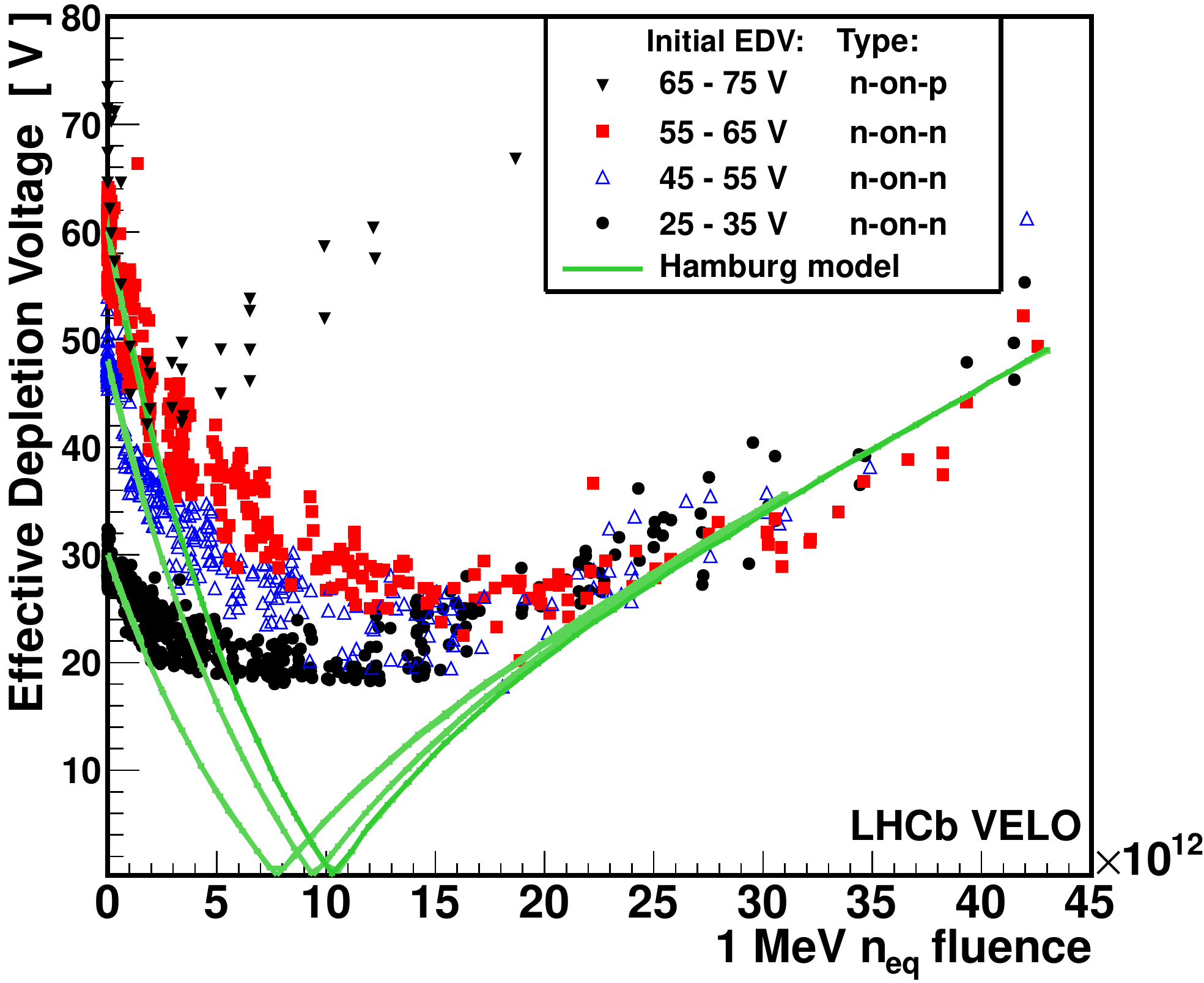}
\caption{The EDV against fluence for VELO sensors of various initial EDV. The
EDV from data is compared to depletion voltages predicted by the Hamburg model,
with good agreement observed prior to, and after sensor type-inversion.}
\label{fig:EDVAll}
\end{figure}

The global change in EDV is determined by combining the data from many of the
VELO sensors with the predicted fluence (see section~\ref{FluDet}), as shown in
figure~\ref{fig:EDVAll}. Sensors are divided into categories based on their
initial EDVs. The irradiation-induced change in the depletion voltage of \nonn{} type sensors is modeled
as a function of time, temperature and fluence by the Hamburg Model. It has
three components: a short term annealing component, a stable damage component
and a reverse annealing component. Taking into account the LHCb luminosity
measurements and VELO sensor temperature readings, the Hamburg model
predictions can be compared to data, as shown by the overlaid curves in
figure~\ref{fig:EDVAll}. Good agreement is found for low fluences, and for
higher fluences after type inversion. It is assumed that the sensors type
invert at a fluence near to the EDV minimum. For all \nonn{} type sensors this
occurs at approximately the same fluence of $\rm{(10-15) \times
10^{12}\,1\,MeV\,n_{eq}}$. The behaviour after inversion is found to be
independent of the initial EDV of the sensor, with an approximately linear
increase in EDV with further fluence.
A linear fit to the data gives
a voltage increase with fluence of $\rm{(1.35 \pm 0.25)
\times 10^{-12}\,V/1\,MeV\,n_{eq}}$. 
For the \nonp{} type, the initial decrease in EDV occurred up to a
fluence of approximately $\rm{2 \times 10^{12}\,1\,MeV\,n_{eq}}$. After this
the EDV has increased with further fluence. The rate of increase is measured to be $\rm{(1.43 \pm 0.16) \, \times 10^{-12}
\,V/1\,MeV\,n_{eq}}$, similar to
that of the type inverted \nonn{} type sensors.

The EDV of the \nonp{} type sensors begins to increase having received
significantly less fluence than the \nonn{} type sensors. If the comparable rate
of EDV increase is maintained with further fluence then the \nonp{} type
sensors will reach an EDV of $500\,\rm{V}$, the hardware limit of the VELO
system, after receiving approximately $\rm{35 \times 10^{12}\,1\,MeV\,n_{eq}}$
less fluence than the \nonn{} type sensors. It is expected that the \nonn{}
sensors will reach the $500\,\rm{V}$ limit following a fluence of approximately
$\rm{380 \times 10^{12}\,1\,MeV\,n_{eq}}$.

The amount of charge collected is expected to change with fluence due to
radiation induced changes to the silicon. 
For $\phi$-type sensors the MPV has decreased by approximately
$4\%$ in the most irradiated regions, having received a fluence of $\rm{40
\times 10^{12}\,1\,MeV\,n_{eq}}$. An even larger reduction of approximately
$8\%$ is found in the inner regions of the R-type sensors, having received a
comparable fluence. This is due to a charge loss mechanism related to the
second metal layer of the R-type sensors, which is described in detail in
section~\ref{sec:secondmetallayer}. The outer regions of the sensor are most
significantly affected, with decreases of approximately $12\%$ observed
following a fluence of just $\rm{2 \times 10^{12}\,1\,MeV\,n_{eq}}$.

\subsection{Noise scans}
\label{sec:noise}
The CCE scan data described in section~\ref{sec:CCE} requires proton beams,
and so is collected at the expense of physics data. A second method has
been developed to monitor radiation damage, using the relationship between
the intrinsic electronic noise of the pre-amplifier and the capacitance of the
sensor. Data scans for this study can be collected regularly as proton
collisions are not required.

In undepleted silicon, several sources of input capacitance are identified, the
most dominant of which is the inter-strip impedance. For \nonn{} sensors before
type inversion, the depletion region grows with increasing voltage from the
backplane (the opposite side to the strips). When the sensor is fully depleted
the space-charge reaches the strips and the inter-strip resistance increases by
several orders of magnitude, resulting in a decrease in sensor
noise\,\cite{CMS_1, CMS_2}. For \nonn{} type sensors following type inversion
and \nonp{} type sensors, the depletion region grows from the strip side of the
silicon. In this situation the strips are immediately isolated at the
application of a bias voltage and the relationship between noise and voltage
cannot be exploited to extract information related to the depletion voltage.

\begin{figure}[b]
\centering
\includegraphics[width=0.5\textwidth]{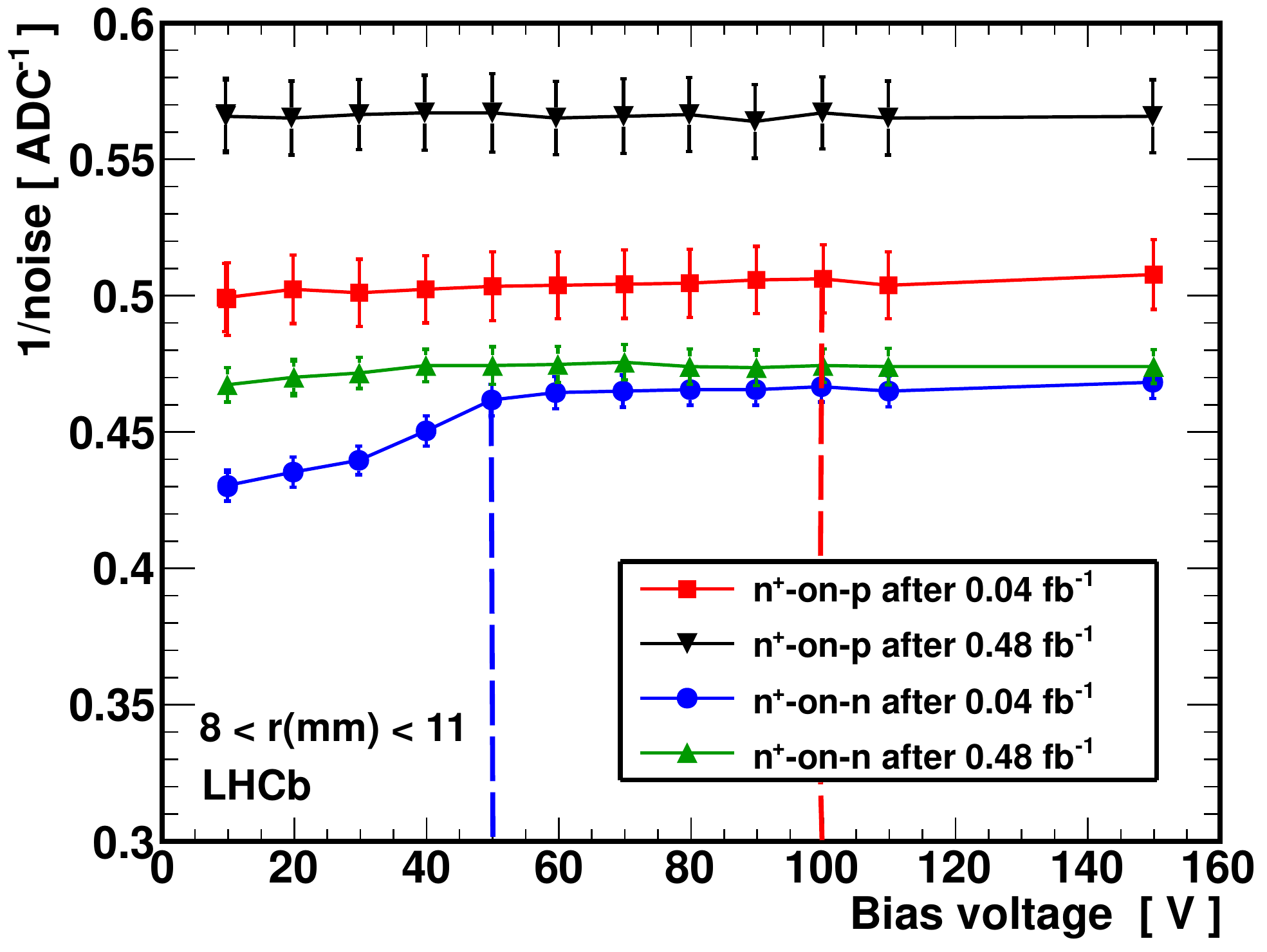}
\caption{The inverse of the sensor noise against bias voltage for a particular
\nonn{} and \nonp{} type sensor, for two values of integrated luminosities. The
C-V scan measured initial depletion voltages are shown by the dashed lines.}
\label{fig:hvscan}
\end{figure}

The intrinsic noise in VELO sensors is determined by subtracting the mean ADC
value (or \emph{pedestal}) and a common-mode noise term. Figure\,\ref{fig:hvscan}
shows the inverse of the intrinsic noise as a function of voltage for an
\nonn{} and \nonp{} type sensor. 
For the \nonp{} before and after
irradiation, and \nonn{} after irradiation (having type inverted) the
distribution is flat, thus little information related to the sensor depletion
voltages can be extracted. For the \nonn{} type sensor prior to type inversion,
an increase in voltage results in a decrease in noise until a plateau is
reached when the sensor is fully depleted.

The noise scan data can be used to identify whether an \nonn{} type sensor
has undergone type inversion. Only R-type sensors are investigated as the strip orientation
allows the identification of strips that have been subject to a specific
fluence. Following a delivered luminosity of approximately
$0.80\,\rm{fb^{-1}}$, $40$ of the sensors are identified as having type-inverted
in the first radial region ($8\hy 11{\,\rm mm}$), while in the second region
($11\hy 16{\,\rm mm}$) $21$ sensors are identified. Similar information can be
extracted from the CCE data, with a sensor region defined as type-inverted when
the measured EDV has reached a minimum and subsequently begun to increase.
Following the same luminosity, the CCE method identified $21$ and $5$
type-inverted sensors in the first and second radial regions, considerably
fewer than the noise method. This discrepancy is understood by examination of
figure~\ref{fig:EDVAll}, in which the minimum of the Hamburg model prediction and
the point at which the EDVs begin to increase are separated by a fluence of
approximately $10\times 10^{12}\,1\,\rm{MeV\,n_{eq}}$. The noise
scan method is not subject to the same fluence lag. Following a delivered
luminosity of $1.2\,\rm{fb^{-1}}$, the CCE method identifies $39$ and $21$
sensors in the two radial regions. This is in good agreement with the noise
method, with the same $39$ and $21$ sensors identified by each method. 

\section{Charge loss to the second metal layer}
\label{sec:secondmetallayer}
\begin{figure}[b]
\centering
\subfigure[]{
      \includegraphics[width=0.47\textwidth]{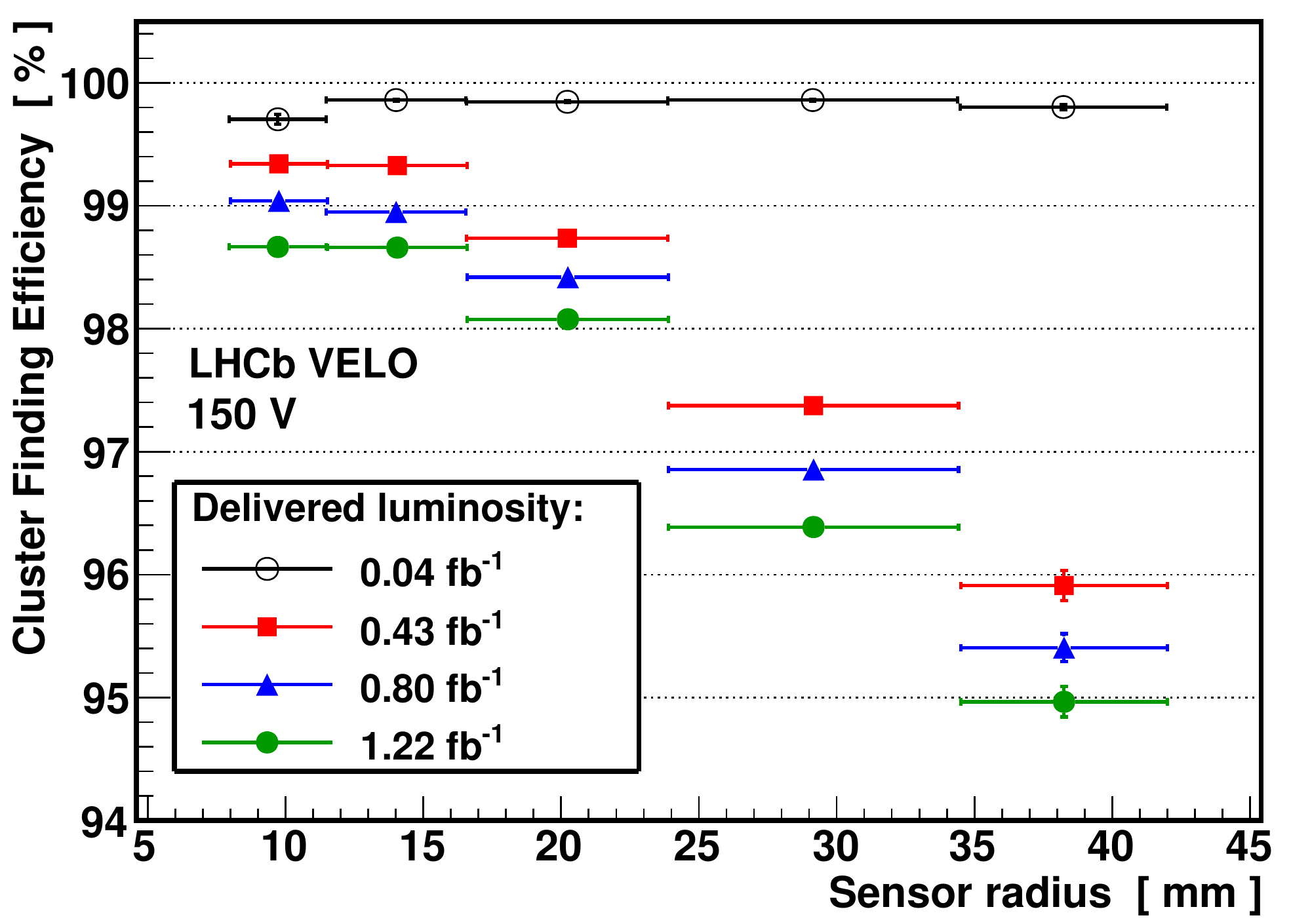}
      \label{fig:CFE_scan}
}
\subfigure[]{
      \includegraphics[width=0.47\textwidth]{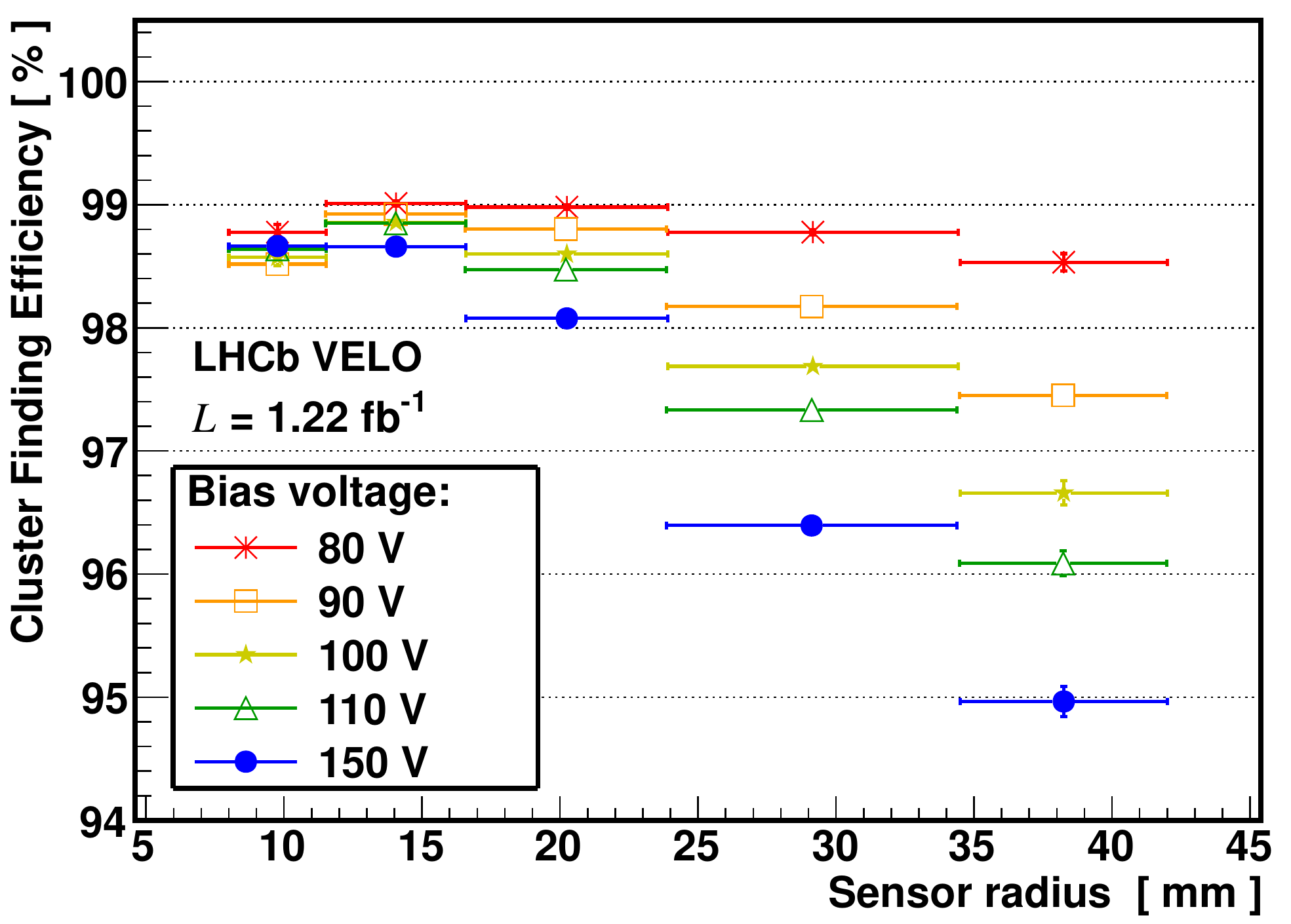}
      \label{fig:CFEVolt}
}
\caption{The CFE for an R-type \nonn{} sensor as a function of sensor radius
for \textbf{a)} different amounts of delivered luminosity and \textbf{b)}
several different bias voltages.}
\label{fig:CFE_drop}
\end{figure}

All physics analyses at LHCb rely on efficient track reconstruction using
clusters from the VELO sensors. A cluster is defined as one or several adjacent
strips with charge above a particular threshold. The data samples described in section~\ref{sec:CCE} are also used to measure the
Cluster Finding Efficiency (CFE), by looking for the presence of a
cluster at the track intercept on the test sensor. Before irradiation the mean
CFE of the VELO sensors was greater than $99\%$ \cite{VELOPP}. After irradiation
the CFE in many sensors decreased significantly, as shown in
figure~\ref{fig:CFE_drop}. The inefficiency is particularly prevalent at large
sensor radii and for high bias voltages. 
The decrease in CFE does not
appear to be proportional to the delivered luminosity, but instead exhibits a
rapid drop between the $0.04$ and $0.43\,\rm{fb^{-1}}$ data scans. 

\begin{figure}[t]
\centering
 \subfigure[]{
  \includegraphics[width=0.45\textwidth]{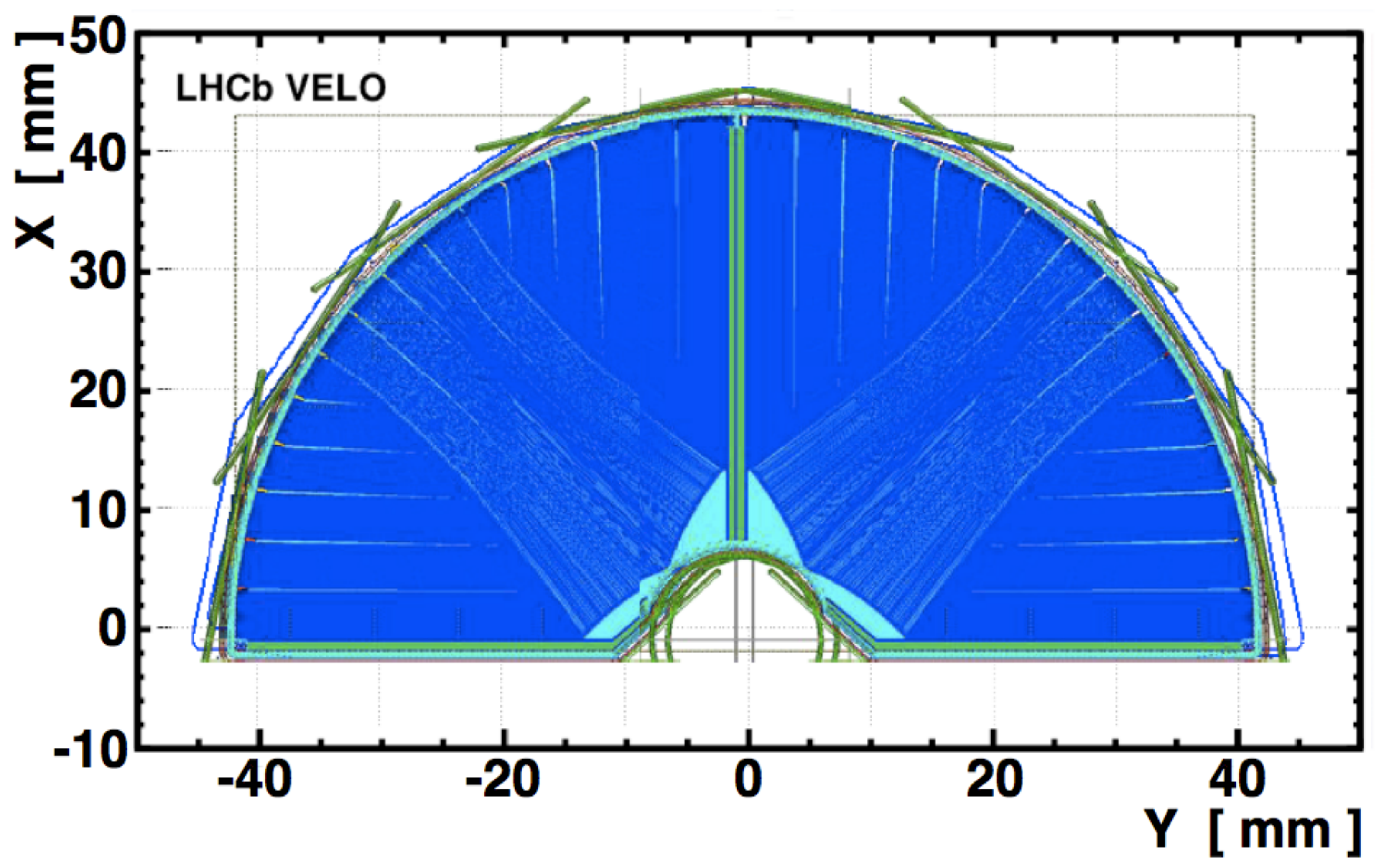}
  \label{fig:CFE2D_A}
 }
\subfigure[]{
  \includegraphics[width=0.505\textwidth]{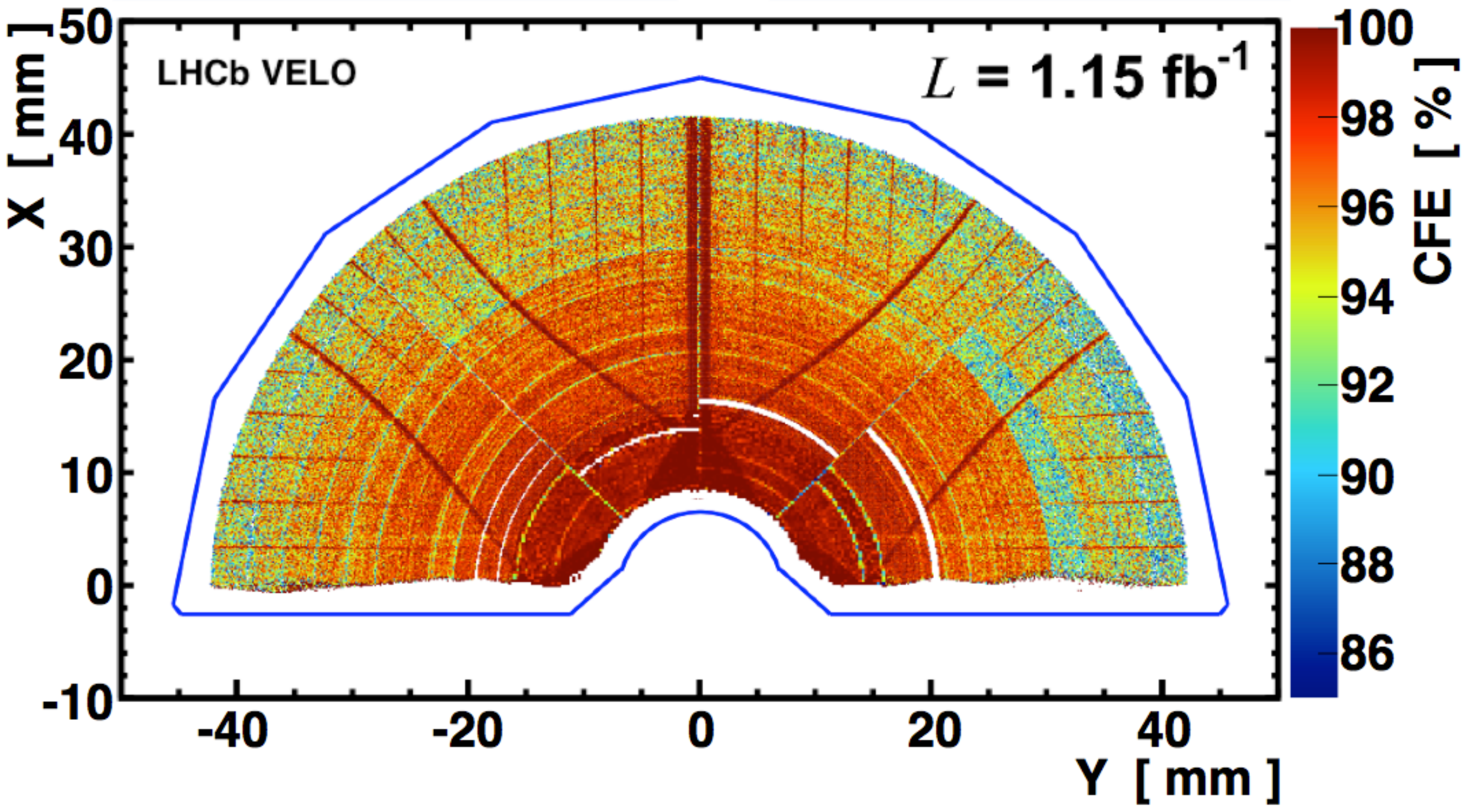}
  \label{fig:CFE2D_B}
 }
\caption{\textbf{a)} The layout of the second metal layer routing lines on an
R-type sensor. The darker regions represent the presence of routing lines, and
the lighter regions their absence. \textbf{b)} The CFE shown in small spatial
regions of an R-type sensor.}
\label{fig:CFE2D}
\end{figure}

\begin{figure}[b]
\centering
\includegraphics[width=0.42\textwidth]{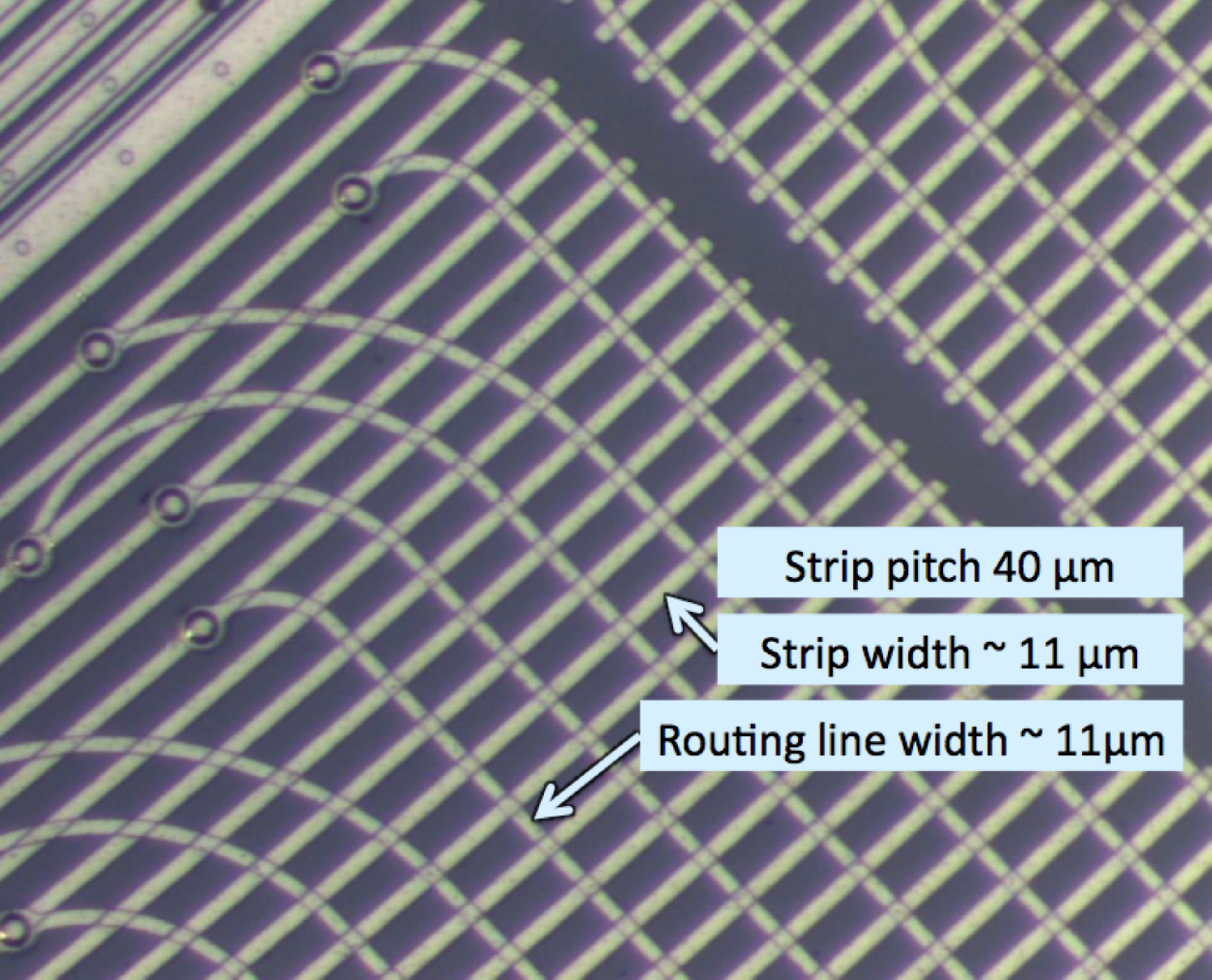}
\caption{A photograph of the innermost region of an R-type sensor. Strips run
from the bottom-left to the top-right. Each strip is connected to a routing
line orientated perpendicularly to the strip.}
\label{fig:RLPhoto}
\end{figure}

\begin{figure}[htbp]
\centering
\includegraphics[width=0.8\textwidth]{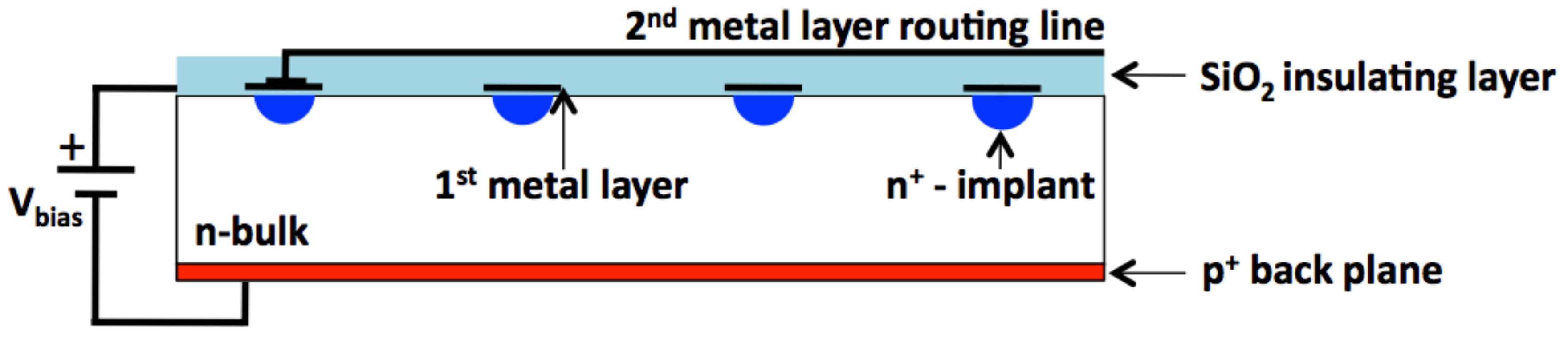}
\caption{A schematic cross-section of a portion of an R-type sensor,
showing the relative position of the two metal layers used to carry the readout
signals in \nonn{} type sensors. The $n^{+}$ implants and strips (into the
page) run perpendicularly to the routing lines (left to right). For clarity the
routing line of just one strip is shown.}
\label{fig:ReadoutRL}
\end{figure}

To determine the source of this CFE decrease, a large sample of regular LHCb
physics data has been used to measure the CFE for small spatial regions on a
sensor at the nominal $\rm{150\,V}$ bias. The result of this is shown in 
figure~\ref{fig:CFE2D}, displayed beside a diagram illustrating the
layout of the second metal layer readout lines. There is a clear correspondence
between the two figures, with high CFE measured in regions that are devoid of
second metal layer lines. The relative orientation of the strips and routing
lines in R-type sensors is shown in figure~\ref{fig:RLPhoto}.
In addition, a schematic cross-section of an R-type sensor is shown in
figure~\ref{fig:ReadoutRL}. The routing lines from inner strips are seen to pass
approximately perpendicularly over the outer strip implants.

\begin{figure}[hbtp]
\centering
\includegraphics[width=0.5\textwidth]{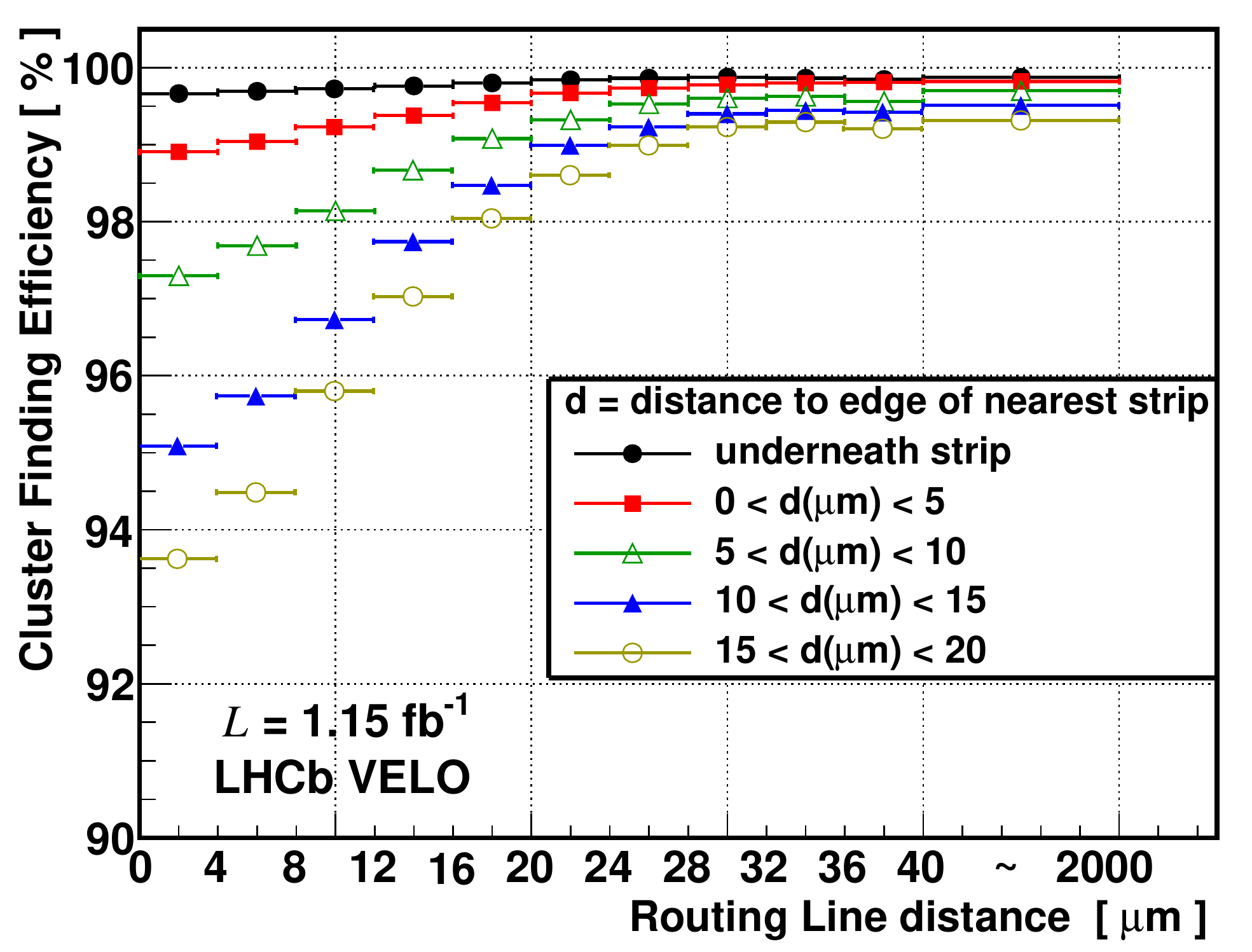}
\caption{The CFE as a function of the distance between the
particle intercept and the nearest routing line, for several bins of the
distance between the particle intercept and closest strip edge.}
\label{fig:RLStripDist}
\end{figure}

Using the precision tracking of the VELO, it is possible to investigate the CFE
loss as a function of the distance between a track intercept with a sensor, and
the nearest strip and routing line. This is shown in
figure~\ref{fig:RLStripDist}. The CFE is improved for track intercepts that are
near to the strip implants. Conversely, the CFE is reduced when a track
intercept is far from a strip and near to a routing line. Similar effects have
been observed in other experiments\,\cite{atlasdm,cmsdm}.

The source of the CFE loss is hypothesised in terms of charge induction on the
second metal layer. Prior to irradiation, ionised electrons will drift along the
field lines, most of which terminate at the $n^{+}$ implants. Hence the
majority of the signal will be induced on the implants, which are strongly
capacitively coupled to the readout strips. The drifting charge is expected to
be collected well within the $\regsim 20$\,ns readout period of the
electronics, and no signal is expected on neighbouring electrodes (with the
exception of capacitive coupling and cross-talk effects, which are measured to
be low).
However, irradiation may cause modifications to the field line structure, such
that not all field lines terminate on the implants. In addition, there may be
charge trapping effects which delay the drift of charge, resulting in
charge sampling before the electrons have reached the implants. In both of these
situations there will be a net induced charge on nearby electrodes, such as the
second metal layer routing lines. In figure~\ref{fig:CFEVolt} the CFE was seen to
worsen with increasing bias voltage. This appears to disfavour the contribution
due to trapping, as an increase in bias voltage should result in faster
collection times. However, the bias voltage may also affect the field-line
structure. In reality, it is likely that the charge loss to the second metal
layer is due to several competing mechanisms.

The CFE loss also exhibits a significant radial dependence, as was shown by
figure~\ref{fig:CFE_drop}. This can be understood by considering two competing
mechanisms. The implant strip width and the fractional area covered by the
strips increases with radius, resulting in reduced charge loss, due to greater
strip shielding. However, the fractional area covered by the second metal layer
also increases with radius, due to the greater density of lines, increasing the
amount of pickup. The latter effect is dominant, hence the overall charge loss
is greater at large sensor radii.

\begin{figure}[t]
\centering
\subfigure[]{
      \includegraphics[width=0.45\textwidth]{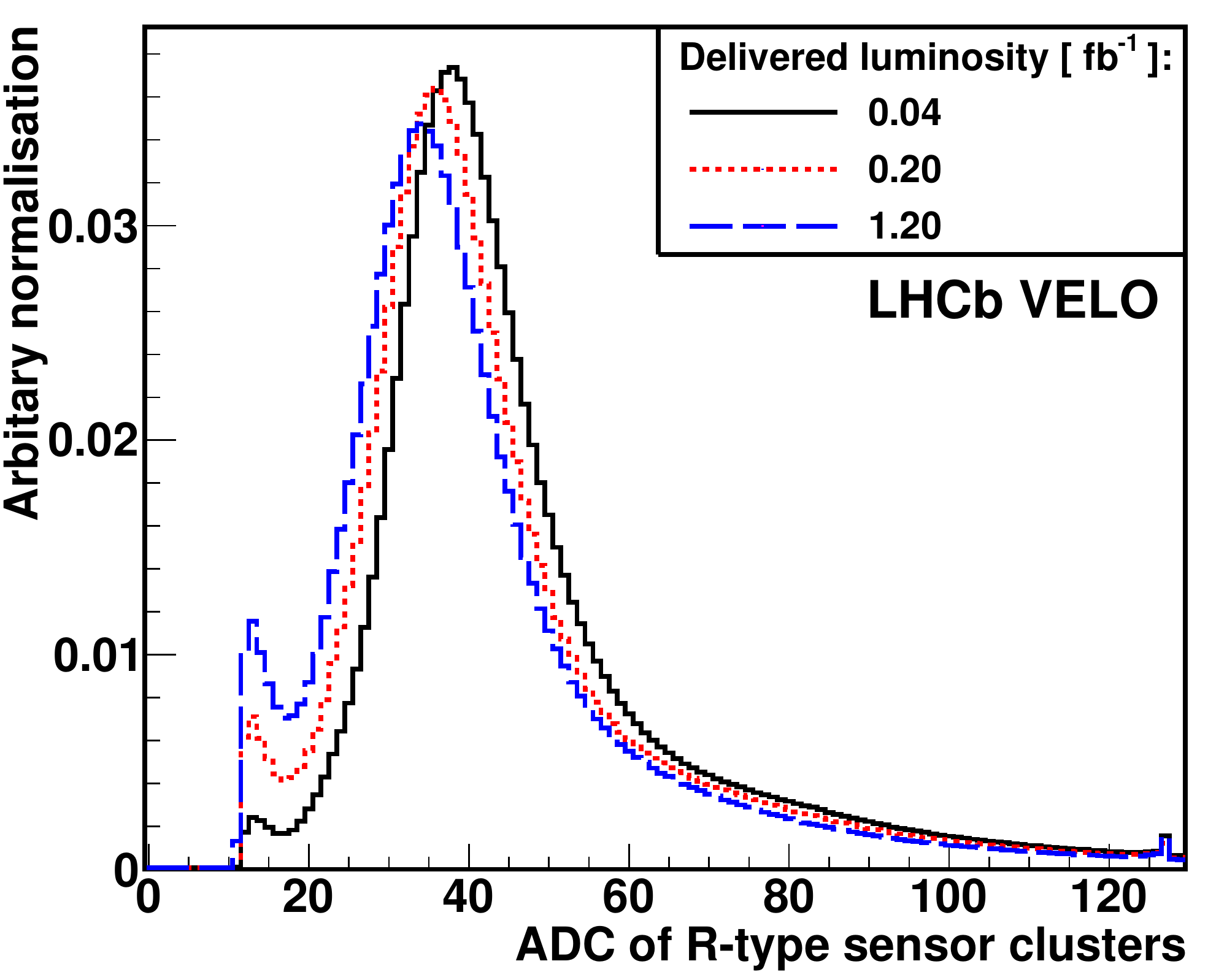}
      \label{fig:NoiseDist}
}
\subfigure[]{
      \includegraphics[width=0.45\textwidth]{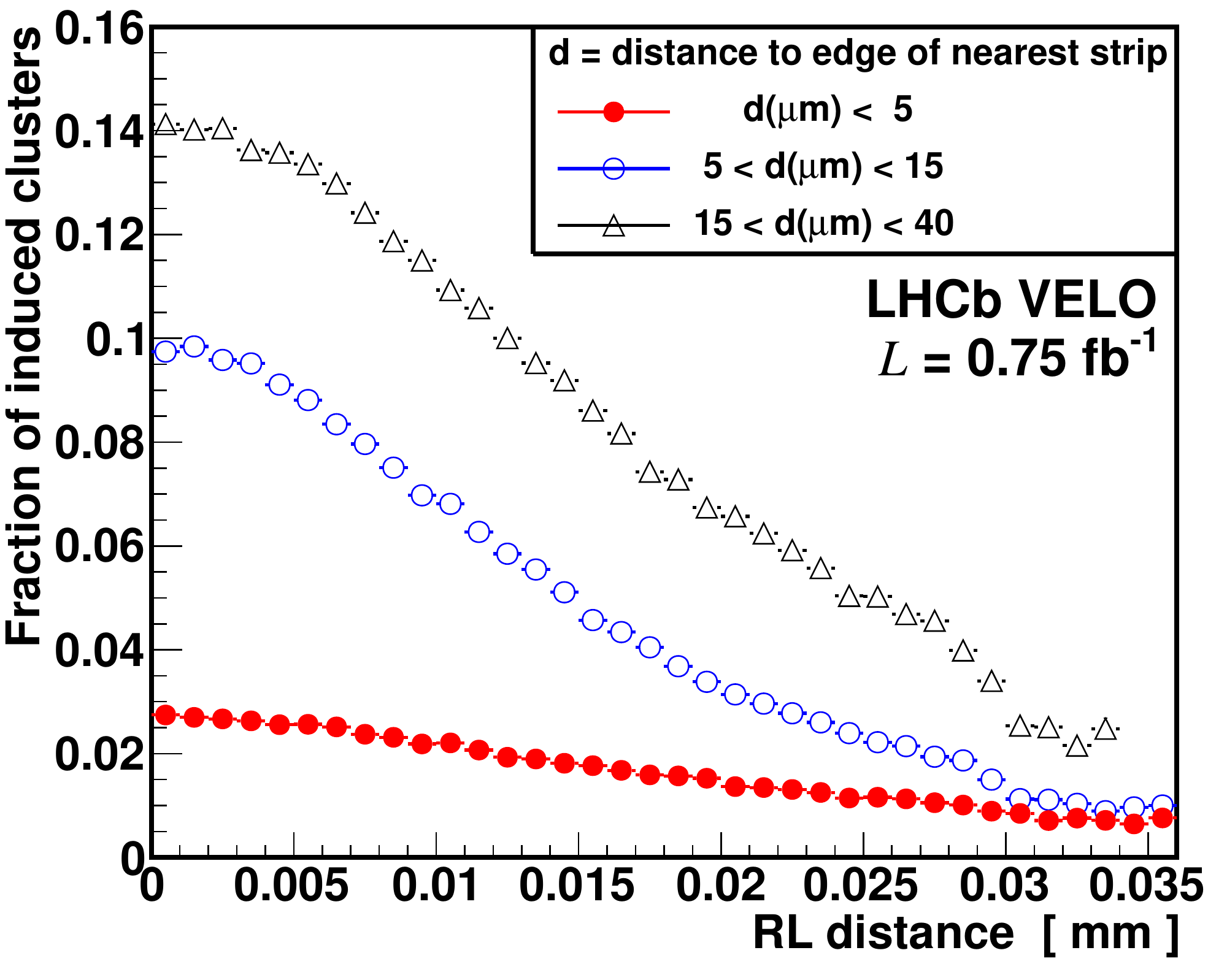}
      \label{fig:FracNoiseClust}
}
\caption{ For R-type sensors: \textbf{a)} The ADC spectrum of all clusters seen
for three different integrated luminosities. The limit at $10$ ADC counts is
imposed by the clustering thresholds. \textbf{b)} The fraction of
reconstructed clusters that are induced on routing lines as a function of the
distance to the nearest routing line and strip. This is determined from the
number of track intercepts for which the inner strip associated to the
nearest routing line has a $1$ strip cluster with less than $35$ ADC counts.}
\label{fig:NoiseClusters}
\end{figure}

In addition to lowering the clustering efficiency, charge induced on a routing
line can introduce a noise cluster. The cluster ADC distribution from R-type
sensors has a peak associated to these low ADC noise clusters that has grown
with fluence, as shown in figure~\ref{fig:NoiseDist}. These noise clusters are
predominantly single strip clusters located at small radius regions of
R-type sensors. The fraction of the induced noise clusters increases when
tracks traverse a sensor near to a routing line and far from a strip, as
shown in figure~\ref{fig:FracNoiseClust}. 

The CFE decrease is not observed in $\phi$-type sensors as the routing
lines from inner strips were intentionally placed directly above the outer
strips to minimise pick-up. This is made possible by the
$\phi$-type sensors strip orientation. 
The CFE loss could be partially
recovered by lowering the cluster reconstruction thresholds. However this comes
at the expense of a worse signal to background ratio which leads to higher
rates of fake tracks reconstructed from noise induced clusters. 

\begin{figure}[t]
\centering
\subfigure[]{
  \includegraphics[width=0.45\textwidth]{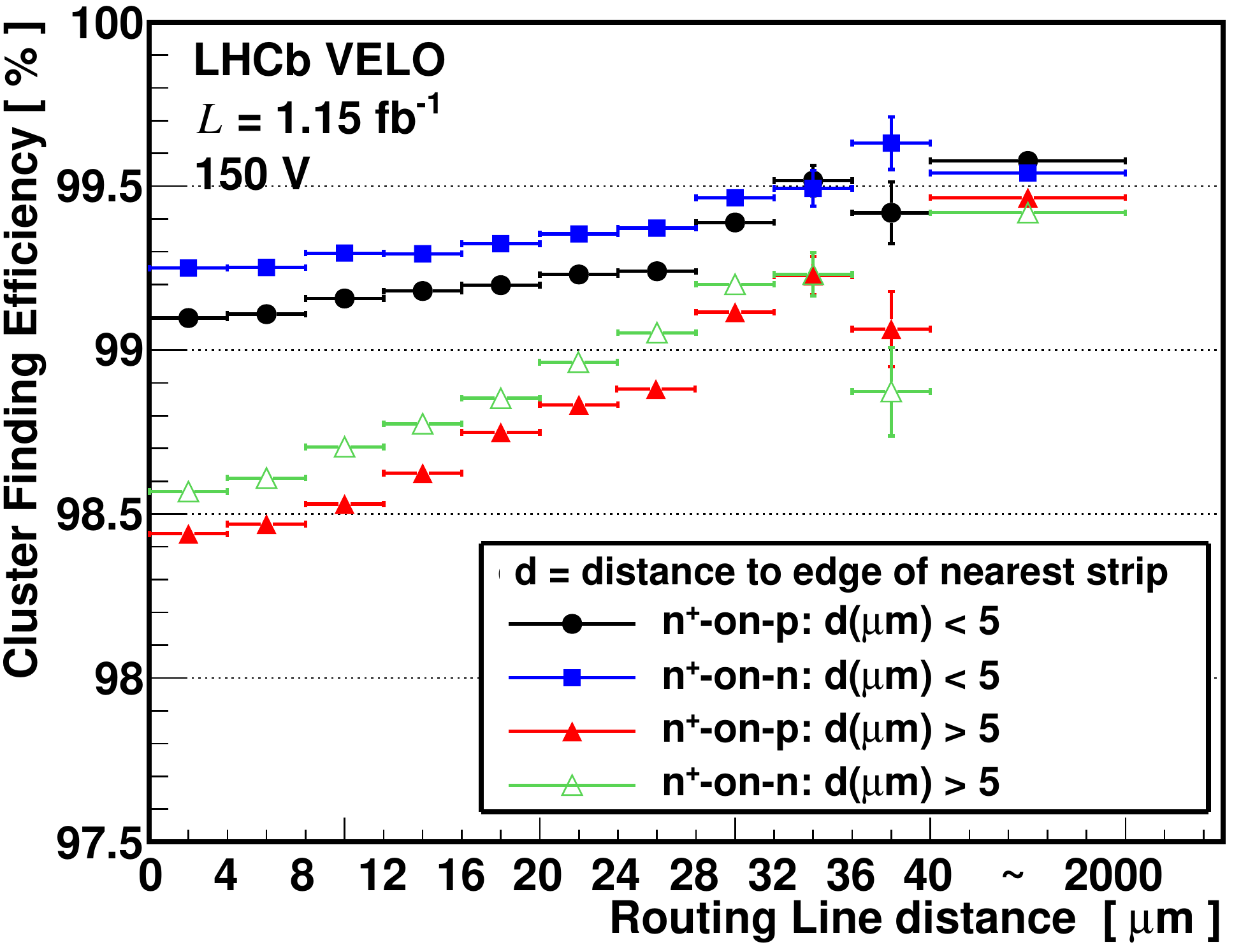}
  \label{fig:RLCFEnonp1}
 }
 \subfigure[]{
  \includegraphics[width=0.45\textwidth]{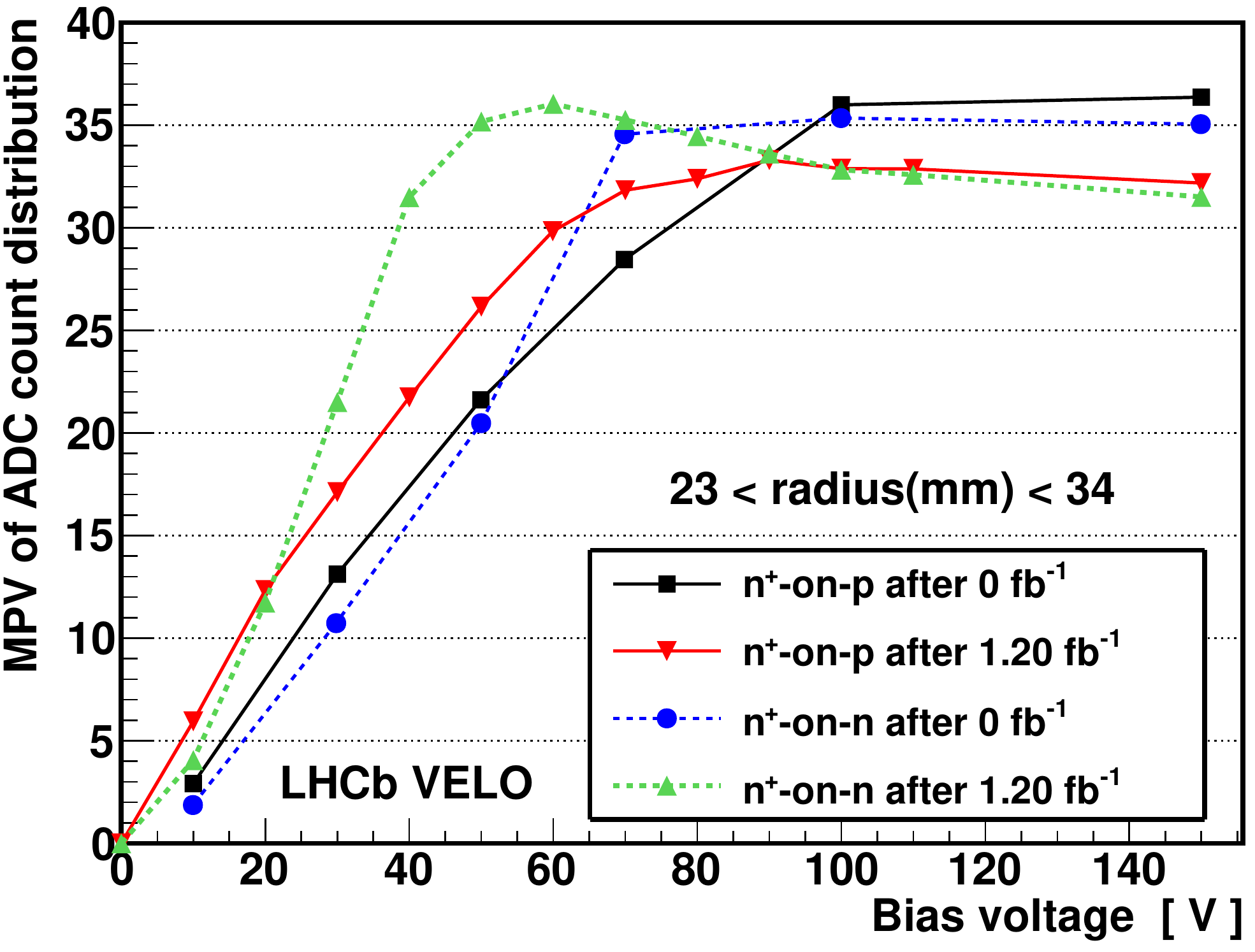}
  \label{fig:RLCFEnonp2}
 }
\caption{\textbf{a)} The CFE as a function of the distance between the
particle intercept and the nearest routing line and strip edge,
compared for an \nonn{} and \nonp{} sensor. \textbf{b)} The MPV vs.
bias voltage for an \nonn{} and \nonp{} sensor.}
\label{fig:RLCFEnonp}
\end{figure}

The R-type \nonp{} and \nonn{} sensors have a similar CFE dependence on
the strip and routing line distance, as shown in figure~\ref{fig:RLCFEnonp1}.
Figure\,\ref{fig:RLCFEnonp2} shows the MPV of the collected charge distribution
as a function of bias voltage for an \nonn{} and an \nonp{} type sensor. At
$150\rm{\,V}$, the MPV of the \nonn{} and \nonp{} are approximately equal, both
before and after irradiation. For the \nonn{} type sensor post irradiation, the
MPV reaches a maximum at around $60\rm{\,V}$ after which it is observed to
decrease with increasing bias voltage. This decrease in MPV leads to a reduced
CFE and is associated with the second metal layer effect. Therefore less charge
is lost to the second metal layer when operating the \nonn{} sensor at a lower
than nominal voltage.

\begin{figure}[t]
\centering
\includegraphics[width=0.5\textwidth]{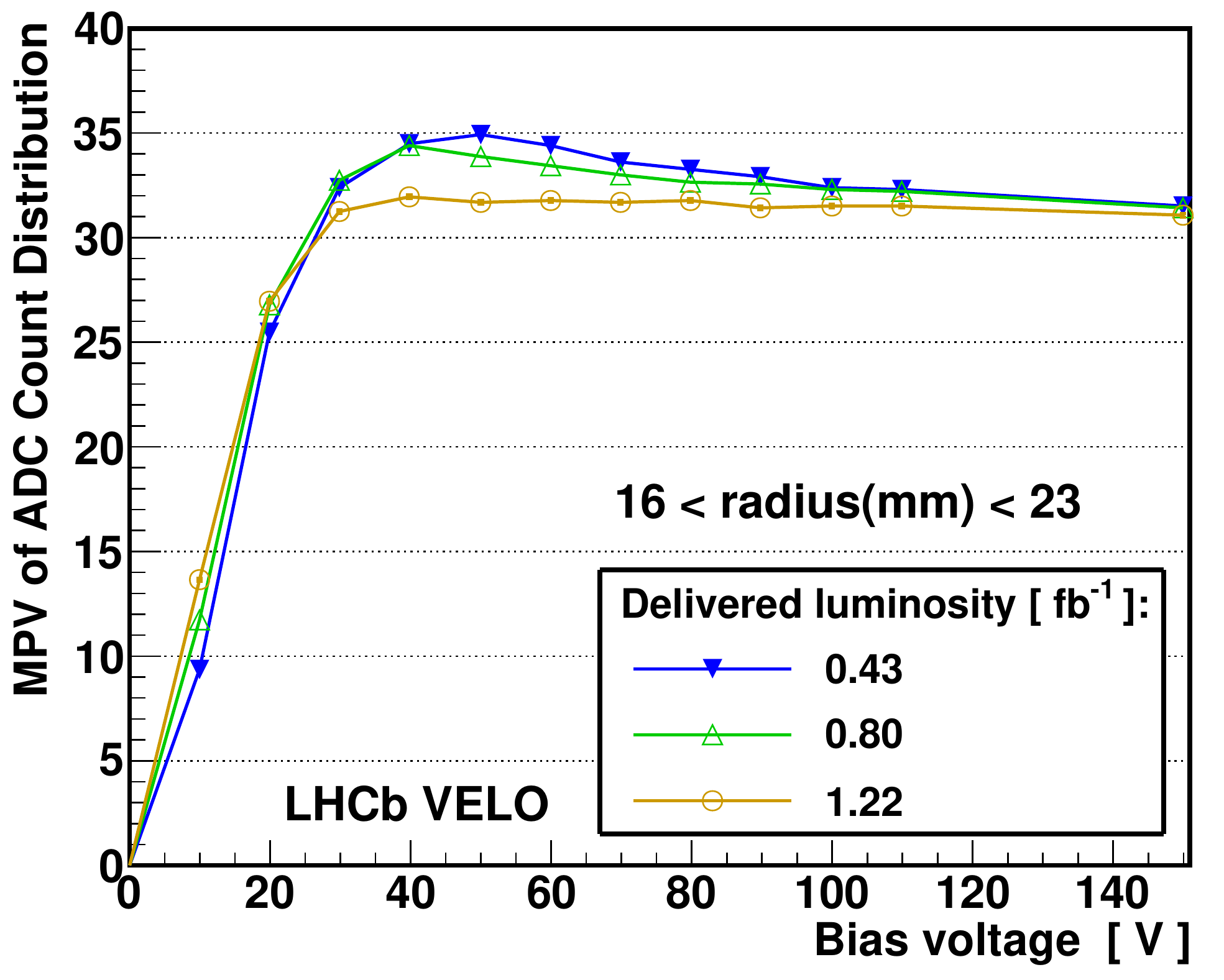}
\caption{The MPV against bias voltage for an R-type sensor at
three values of delivered luminosity. The sensor region has been identified
as having type inverted in the $1.20\,\rm{fb^{-1}}$ scan, in which the MPV
dependence on voltage is no longer present.}
\label{fig:S23_MPV_voltWithFlu}
\end{figure}

The \nonp{} type sensor does not exhibit this same charge collection loss (and
resulting CFE decrease) dependence on voltage. This may be due to the depletion
region in the \nonp{} type sensor growing from the strip side of the silicon
instead of from the sensor backplane. This is supported by the observation that
after type inversion, charge loss in \nonn{} type sensors
no longer depends on the bias voltage, as shown in
figure~\ref{fig:S23_MPV_voltWithFlu}. 

\section{Summary}
\label{sec:Summary}
The effects of radiation damage have been observed in all of the LHCb
VELO sensors. At the operational sensor temperature of approximately
$-7\,^{\circ}\rm{C}$, the average rate of sensor current increase is measured to be
$18\,\upmu\rm{A}$ per $\rm{fb^{-1}}$, in agreement with expectations. The silicon
effective bandgap has been determined using current versus temperature scan
data collected at various levels of radiation, with an average value of
$E_{g}=1.16\pm0.03\pm0.04\,\rm{eV}$ found. 

Analysis of the Charge Collection Efficiency has proven an effective method for
tracking the evolution of the sensor depletion voltages with fluence.
Measurements of the Effective Depletion Voltage after sensor type inversion have
been shown to agree well with the Hamburg model predictions. A method
relating the sensor noise dependence on bias voltage was used to
identify type-inverted sensors, and was found to be in good agreement with the
Charge Collection Efficiency method.

A significant decrease in the Cluster Finding Efficiency in R-type
sensors due to the second metal layer has been observed following a relatively
small amount of particle fluence. In the worst affected sensor regions the
Cluster Finding Efficiency decreased by over $5\%$. Despite this relatively
large localised inefficiency, studies of the VELO tracking efficiencies show no
degradation associated to this effect, within the errors of $\pm 0.3\%$. For
the \nonn{} type sensors before type inversion the magnitude of the charge loss
is found to increase with sensor bias voltage. For type-inverted \nonn{} type
sensors and \nonp{} type sensors, a voltage dependence is not observed.
Radiation induced modifications to the field line structure and charge trapping effects are thought to be possible sources of this charge loss effect. 

The two \nonp{} sensors have been studied in detail, providing
valuable information for upgraded detector designs. If the Effective Depletion
Voltage were to continue increasing at the currently observed rate with further
irradiation, then the \nonp{} would reach the $500$\,V hardware limit having
received approximately $\rm{35 \times 10^{12}\,1\,MeV\,n_{eq}}$ less fluence
than an equivalent \nonn{} type sensor. This corresponds to approximately
$1\,\rm{fb}^{-1}$ of integrated luminosity in the highest particle fluence
region of the VELO.

\section*{Acknowledgements}

\noindent We express our gratitude to our colleagues in the CERN
accelerator departments for the excellent performance of the LHC. We
thank the technical and administrative staff at the LHCb
institutes. We acknowledge support from CERN and from the national
agencies: CAPES, CNPq, FAPERJ and FINEP (Brazil); NSFC (China);
CNRS/IN2P3 and Region Auvergne (France); BMBF, DFG, HGF and MPG
(Germany); SFI (Ireland); INFN (Italy); FOM and NWO (The Netherlands);
SCSR (Poland); ANCS/IFA (Romania); MinES, Rosatom, RFBR and NRC
``Kurchatov Institute'' (Russia); MinECo, XuntaGal and GENCAT (Spain);
SNSF and SER (Switzerland); NAS Ukraine (Ukraine); STFC (United
Kingdom); NSF (USA). We also acknowledge the support received from the
ERC under FP7. The Tier1 computing centres are supported by IN2P3
(France), KIT and BMBF (Germany), INFN (Italy), NWO and SURF (The
Netherlands), PIC (Spain), GridPP (United Kingdom). We are thankful
for the computing resources put at our disposal by Yandex LLC
(Russia), as well as to the communities behind the multiple open
source software packages that we depend on. A special acknowledgement goes to
all our LHCb collaborators who over the years have contributed to obtain the
results presented in this paper.

\bibliographystyle{LHCb}
\bibliography{main2}

\ifx\mcitethebibliography\mciteundefinedmacro
\PackageError{LHCb.bst}{mciteplus.sty has not been loaded}
{This bibstyle requires the use of the mciteplus package.}\fi
\providecommand{\href}[2]{#2}
\begin{mcitethebibliography}{10}
\mciteSetBstSublistMode{n}
\mciteSetBstMaxWidthForm{subitem}{\alph{mcitesubitemcount})}
\mciteSetBstSublistLabelBeginEnd{\mcitemaxwidthsubitemform\space}
{\relax}{\relax}

\bibitem{LHCbDet}
LHCb collaboration, A.~A. Alves~Jr. {\em et~al.},
  \ifthenelse{\boolean{articletitles}}{{\it {The \lhcb detector at the LHC}},
  }{}\href{http://dx.doi.org/10.1088/1748-0221/3/08/S08005}{JINST {\bf 3}
  (2008) S08005}\relax
\mciteBstWouldAddEndPuncttrue
\mciteSetBstMidEndSepPunct{\mcitedefaultmidpunct}
{\mcitedefaultendpunct}{\mcitedefaultseppunct}\relax
\EndOfBibitem
\bibitem{Moll}
M.~Moll, {\em \href{https://mmoll.web.cern.ch/mmoll/thesis/}{Radiation Damage
  in Silicon Particle Detectors - microscopic defects and macroscopic
  properties}}, PhD thesis, Fachbereich Physik der Universit\"{a}t Hamburg,
  1999\relax
\mciteBstWouldAddEndPuncttrue
\mciteSetBstMidEndSepPunct{\mcitedefaultmidpunct}
{\mcitedefaultendpunct}{\mcitedefaultseppunct}\relax
\EndOfBibitem
\bibitem{TempDepCur}
Y.~Wang, A.~Neugroschel, and C.~T. Sah,
  \ifthenelse{\boolean{articletitles}}{{\it {Temperature Dependence of Surface
  Recombination Current in {MOS} Transistors}},
  }{}\href{http://dx.doi.org/10.1109/16.944201}{IEEE Trans.\ Nucl.\ Sci.\  {\bf
  48} (2001) 2095}\relax
\mciteBstWouldAddEndPuncttrue
\mciteSetBstMidEndSepPunct{\mcitedefaultmidpunct}
{\mcitedefaultendpunct}{\mcitedefaultseppunct}\relax
\EndOfBibitem
\bibitem{IVNote}
A.~Gureja {\em et~al.}, \ifthenelse{\boolean{articletitles}}{{\it Use of {IV}
  (current vs voltage) scans to track radiation damage in the {LHCb} {VELO}},
  }{}
  \href{http://cdsweb.cern.ch/search?p=LHCb-PUB-2011-020&f=reportnumber&action%
_search=Search&c=LHCb+Reports&c=LHCb+Conference+Proceedings&c=LHCb+Conference+%
Contributions&c=LHCb+Notes&c=LHCb+Theses&c=LHCb+Papers}
  {LHCb-PUB-2011-020}\relax
\mciteBstWouldAddEndPuncttrue
\mciteSetBstMidEndSepPunct{\mcitedefaultmidpunct}
{\mcitedefaultendpunct}{\mcitedefaultseppunct}\relax
\EndOfBibitem
\bibitem{ITNote}
A.~Hickling {\em et~al.}, \ifthenelse{\boolean{articletitles}}{{\it Use of {IT}
  (current vs temperature) scans to study radiation damage in the {LHCb}
  {VELO}}, }{}
  \href{http://cdsweb.cern.ch/search?p=LHCb-PUB-2011-021&f=reportnumber&action%
_search=Search&c=LHCb+Reports&c=LHCb+Conference+Proceedings&c=LHCb+Conference+%
Contributions&c=LHCb+Notes&c=LHCb+Theses&c=LHCb+Papers}
  {LHCb-PUB-2011-021}\relax
\mciteBstWouldAddEndPuncttrue
\mciteSetBstMidEndSepPunct{\mcitedefaultmidpunct}
{\mcitedefaultendpunct}{\mcitedefaultseppunct}\relax
\EndOfBibitem
\bibitem{EgGap}
A.~Chilingarov, \ifthenelse{\boolean{articletitles}}{{\it Generation current
  temperature scaling}, }{} Tech. Rep. PH-EP-Tech-Note-2013-001, CERN, Geneva,
  Jan, 2013\relax
\mciteBstWouldAddEndPuncttrue
\mciteSetBstMidEndSepPunct{\mcitedefaultmidpunct}
{\mcitedefaultendpunct}{\mcitedefaultseppunct}\relax
\EndOfBibitem
\bibitem{ActEnergy}
E.~Fretwurst {\em et~al.}, \ifthenelse{\boolean{articletitles}}{{\it Reverse
  annealing of the effective impurity concentration and long term operational
  scenario for silicon detectors in future collider experiments},
  }{}\href{http://dx.doi.org/10.1016/0168-9002(94)91417-6}{Nucl.\ Instrum.\
  Meth.\  {\bf A342} (1994), no.~1 119}\relax
\mciteBstWouldAddEndPuncttrue
\mciteSetBstMidEndSepPunct{\mcitedefaultmidpunct}
{\mcitedefaultendpunct}{\mcitedefaultseppunct}\relax
\EndOfBibitem
\bibitem{NIEL}
A.~Vasilescu, \ifthenelse{\boolean{articletitles}}{{\it The {NIEL} hypothesis
  applied to neutron spectra of irradiation facilities and in the {ATLAS} and
  {CMS} {SCT}}, }{}
  \href{http://rd48.web.cern.ch/rd48/technical-notes/rosetn.htm#1997}{ROSE/TN/%
97-2}, December, 1999\relax
\mciteBstWouldAddEndPuncttrue
\mciteSetBstMidEndSepPunct{\mcitedefaultmidpunct}
{\mcitedefaultendpunct}{\mcitedefaultseppunct}\relax
\EndOfBibitem
\bibitem{DamFact}
A.~Vasilescu and G.~Lindstroem, \ifthenelse{\boolean{articletitles}}{{\it
  Displacement damage in silicon, on-line compilation}, }{} tech. rep.,
  \href{http://sesam.desy.de/members/gunnar/Si-dfuncs.html}{http://sesam.desy.%
de/members/gunnar/Si-dfuncs.html}\relax
\mciteBstWouldAddEndPuncttrue
\mciteSetBstMidEndSepPunct{\mcitedefaultmidpunct}
{\mcitedefaultendpunct}{\mcitedefaultseppunct}\relax
\EndOfBibitem
\bibitem{Hamburg}
R.~Wunstorf {\em et~al.}, \ifthenelse{\boolean{articletitles}}{{\it Results on
  radiation hardness of silicon detectors up to neutron fluences of
  $\rm{10^{15}~n/cm^{2}}$},
  }{}\href{http://dx.doi.org/10.1016/0168-9002(92)90696-2}{Nucl.\ Instrum.\
  Meth.\  {\bf A315} (1992), no.~1â€“3 149 }\relax
\mciteBstWouldAddEndPuncttrue
\mciteSetBstMidEndSepPunct{\mcitedefaultmidpunct}
{\mcitedefaultendpunct}{\mcitedefaultseppunct}\relax
\EndOfBibitem
\bibitem{nonpPred}
F.~Lemeilleur {\em et~al.}, \ifthenelse{\boolean{articletitles}}{{\it
  Electrical properties and charge collection efficiency for neutron-irradiated
  p-type and n-type silicon detectors},
  }{}\href{http://dx.doi.org/10.1016/0920-5632(93)90054-A}{Nuclear Physics B -
  Proceedings Supplements {\bf 32} (1993), no.~0 415}\relax
\mciteBstWouldAddEndPuncttrue
\mciteSetBstMidEndSepPunct{\mcitedefaultmidpunct}
{\mcitedefaultendpunct}{\mcitedefaultseppunct}\relax
\EndOfBibitem
\bibitem{SensComp}
M.~Lozano {\em et~al.}, \ifthenelse{\boolean{articletitles}}{{\it Comparison of
  radiation hardness {P}-in-{N}, {N}-in-{N}, {N}-in-{P} {S}ilicon {P}ad
  {D}etectors}, }{}\href{http://dx.doi.org/10.1109/TNS.2005.855809}{IEEE
  Trans.\ Nucl.\ Sci.\  {\bf 52} (2005) 1468}\relax
\mciteBstWouldAddEndPuncttrue
\mciteSetBstMidEndSepPunct{\mcitedefaultmidpunct}
{\mcitedefaultendpunct}{\mcitedefaultseppunct}\relax
\EndOfBibitem
\bibitem{LiverpoolDV}
P.~R. Turner, \ifthenelse{\boolean{articletitles}}{{\it {VELO} module
  production - sensor testing}, }{}
  \href{http://cdsweb.cern.ch/search?p=LHCb-2007-072&f=reportnumber&action_sea%
rch=Search&c=LHCb+Reports&c=LHCb+Conference+Proceedings&c=LHCb+Conference+Cont%
ributions&c=LHCb+Notes&c=LHCb+Theses&c=LHCb+Papers} {LHCb-2007-072}\relax
\mciteBstWouldAddEndPuncttrue
\mciteSetBstMidEndSepPunct{\mcitedefaultmidpunct}
{\mcitedefaultendpunct}{\mcitedefaultseppunct}\relax
\EndOfBibitem
\bibitem{CMS_1}
The CMS Tracker Collaboration, C.~Barth,
  \ifthenelse{\boolean{articletitles}}{{\it Evolution of silicon sensor
  characteristics of the {CMS} tracker},
  }{}\href{http://dx.doi.org/10.1016/j.nima.2011.05.045}{Nucl.\ Instrum.\
  Meth.\  {\bf A658} (2011), no.~1 6}\relax
\mciteBstWouldAddEndPuncttrue
\mciteSetBstMidEndSepPunct{\mcitedefaultmidpunct}
{\mcitedefaultendpunct}{\mcitedefaultseppunct}\relax
\EndOfBibitem
\bibitem{CMS_2}
The CMS Tracker Collaboration, C.~Barth,
  \ifthenelse{\boolean{articletitles}}{{\it Evolution of silicon sensor
  characteristics of the {CMS} silicon strip tracker},
  }{}\href{http://dx.doi.org/10.1016/j.nima.2012.05.058}{Nucl.\ Instrum.\
  Meth.\  {\bf In Press} (2012), no.~0 }\relax
\mciteBstWouldAddEndPuncttrue
\mciteSetBstMidEndSepPunct{\mcitedefaultmidpunct}
{\mcitedefaultendpunct}{\mcitedefaultseppunct}\relax
\EndOfBibitem
\bibitem{VELOPP}
The LHCb VELO Group, \ifthenelse{\boolean{articletitles}}{{\it {Perfomance of
  the {LHCb} {V}ertex {L}ocator}}, }{}To be submitted to JINST (2012)\relax
\mciteBstWouldAddEndPuncttrue
\mciteSetBstMidEndSepPunct{\mcitedefaultmidpunct}
{\mcitedefaultendpunct}{\mcitedefaultseppunct}\relax
\EndOfBibitem
\bibitem{atlasdm}
ATLAS Pixel Collaboration, L.~Tommaso,
  \ifthenelse{\boolean{articletitles}}{{\it {Test beam results of ATLAS Pixel
  sensors}}, }{}\href{http://arxiv.org/abs/hep-ex/0210045}{{\tt
  arXiv:hep-ex/0210045}}\relax
\mciteBstWouldAddEndPuncttrue
\mciteSetBstMidEndSepPunct{\mcitedefaultmidpunct}
{\mcitedefaultendpunct}{\mcitedefaultseppunct}\relax
\EndOfBibitem
\bibitem{cmsdm}
T.~Rohe {\em et~al.}, \ifthenelse{\boolean{articletitles}}{{\it {Position
  Dependence of Charge Collection in Prototype Sensors for the CMS Pixel
  Detector}}, }{}\href{http://dx.doi.org/10.1109/TNS.2004.829487}{IEEE Trans.\
  Nucl.\ Sci.\  {\bf 51} (2004) 1150}\relax
\mciteBstWouldAddEndPuncttrue
\mciteSetBstMidEndSepPunct{\mcitedefaultmidpunct}
{\mcitedefaultendpunct}{\mcitedefaultseppunct}\relax
\EndOfBibitem
\end{mcitethebibliography}

\end{document}